\documentclass{article}

\usepackage{arxiv}

\usepackage[utf8]{inputenc} 
\usepackage[T1]{fontenc}    
\usepackage{hyperref}       
\usepackage{url}            
\usepackage{booktabs}       
\usepackage{amsfonts}       
\usepackage{nicefrac}       
\usepackage{microtype}      
\usepackage{lipsum}		
\usepackage{graphicx}
\usepackage[sort&compress,numbers]{natbib}
\usepackage{doi}
\usepackage{tikz}
\usetikzlibrary{decorations.markings}
\usetikzlibrary{decorations.pathreplacing}
\usetikzlibrary{shapes.geometric}
\usepackage{mathtools}
\usepackage{float}
\usepackage{amsmath,amssymb,bm}
\usepackage{amsthm}
\usepackage{multirow}
\usepackage{caption}
\usepackage{booktabs}
\usepackage{siunitx}

\usepackage{mathrsfs}
\usepackage{pgfplots}
\usepackage{caption}
\usepackage[linesnumbered,ruled,vlined]{algorithm2e}
\graphicspath{ {Images/} }
\usepackage{multirow}
\usepackage{makecell}
\usepackage{float}
\usepackage[section]{placeins}
\usepackage[multiple]{footmisc}
\usepackage{hyperref}
\usepackage{braket}
\hypersetup{
    colorlinks=true,
    linkcolor=blue,
    filecolor=magenta,
    urlcolor=cyan,
}
\usepackage{xcolor}

\title{Physics-Informed Neural Networks for an optimal counterdiabatic quantum computation}

\date{\footnotesize{\today}}	

\author{ \href{https://orcid.org/0000-0002-2305-5261}{\includegraphics[scale=0.06]{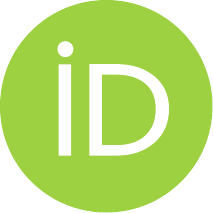}\hspace{1mm}\footnotesize{Antonio Ferrer-Sánchez}}\thanks{\scriptsize{IDAL, Electronic Engineering Department, ETSE-UV, University of Valencia, Avgda. Universitat s/n, 46100 Burjassot, Valencia, Spain.}}\textsuperscript{\textnormal{,}}\thanks{\scriptsize{Valencian Graduate School and Research Network of Artificial Intelligence (ValgrAI), Spain.}} \\
	\footnotesize{\texttt{Antonio.Ferrer-Sanchez@uv.es}}\\
	\footnotesize{\texttt{Corresponding author}}
	\And
	\href{https://orcid.org/0000-0003-4225-6082}{\includegraphics[scale=0.06]{orcid.pdf}\hspace{1mm}\footnotesize{Carlos Flores-Garrigos}}\textsuperscript{\textnormal{1}}\\
	\footnotesize{\texttt{Carlos.Flores-Garrigos@uv.es}} \\
	\And
	\href{https://orcid.org/0000-0001-6893-5660}{\includegraphics[scale=0.06]{orcid.pdf}\hspace{1mm}\footnotesize{Carlos Hernani-Morales}}\textsuperscript{\textnormal{1}}\textsuperscript{\textnormal{,}}\thanks{\scriptsize{Quantum Spain, 46100 Burjassot, Valencia, Spain.}}\\
 \footnotesize{\texttt{Carlos.Hernani@uv.es}} \\
	\And
	\href{https://orcid.org/0009-0004-5826-0910}{\includegraphics[scale=0.06]{orcid.pdf}\hspace{1mm}\footnotesize{José J. Orquín-Marqués}}\textsuperscript{\textnormal{1}}\\
 \footnotesize{\texttt{jose.j.orquin@uv.es}} \\
	\And
	\href{https://orcid.org/0000-0002-9673-2833}{\includegraphics[scale=0.06]{orcid.pdf}\hspace{1mm}\footnotesize{Narendra N. Hegade}}\thanks{\scriptsize{Kipu Quantum, Greifswalderstrasse 226, 10405 Berlin, Germany.}}
	\And
	\href{https://orcid.org/0000-0000-0000-0000}{\includegraphics[scale=0.06]{orcid.pdf}\hspace{1mm}\footnotesize{Alejandro Gomez Cadavid}}\textsuperscript{\textnormal{4}}
 \And
	\href{https://orcid.org/0000-0000-0000-0000}{\includegraphics[scale=0.06]{orcid.pdf}\hspace{1mm}\footnotesize{Iraitz Montalban}}\textsuperscript{\textnormal{4}}
	\And
	\href{https://orcid.org/0000-0002-8602-1181}{\includegraphics[scale=0.06]{orcid.pdf}\hspace{1mm}\footnotesize{Enrique Solano}}\textsuperscript{\textnormal{4}}
	\And
	\href{https://orcid.org/0000-0000-0000-0000}{\includegraphics[scale=0.06]{orcid.pdf}\hspace{1mm}\footnotesize{Yolanda Vives-Gilabert}}\textsuperscript{\textnormal{1}}\\
 \footnotesize{\texttt{yolanda.vives@uv.es}}\\
	\And
	\href{https://orcid.org/0000-0001-9378-0285}{\includegraphics[scale=0.06]{orcid.pdf}\hspace{1mm}\footnotesize{José D. Martín-Guerrero}}\textsuperscript{\textnormal{1,2,3}}\\
 \footnotesize{\texttt{jose.d.martin@uv.es}} 
}

\renewcommand{\headeright}{}
\renewcommand{\undertitle}{A Preprint}

\hypersetup{
pdftitle={QuPINNs: Physics-Informed Neural Networks for an optimal counterdiabatic quantum computation},
pdfkeywords={keywords},
}

\begin{document}
\maketitle

\begin{abstract}
We introduce a novel methodology that leverages the strength of Physics-Informed Neural Networks (PINNs) to address the counterdiabatic (CD) protocol in the optimization of quantum circuits comprised of systems with $N_{Q}$ qubits. The primary objective is to utilize physics-inspired deep learning techniques to accurately solve the time evolution of the different physical observables within the quantum system. To accomplish this objective, we embed the necessary physical information into an underlying neural network to effectively tackle the problem. In particular, we impose the hermiticity condition on all physical observables and make use of the principle of least action, guaranteeing the acquisition of the most appropriate counterdiabatic terms based on the underlying physics. The proposed approach offers a dependable alternative to address the CD driving problem, free from the constraints typically encountered in previous methodologies relying on classical numerical approximations. Our method provides a general framework to obtain optimal results from the physical observables relevant to the problem, including the external parameterization in time known as scheduling function, the gauge potential or operator involving the non-adiabatic terms, as well as the temporal evolution of the energy levels of the system, among others. The main applications of this methodology have been the $\mathrm{H_{2}}$ and $\mathrm{LiH}$ molecules, represented by a 2-qubit and 4-qubit systems employing the STO-3G basis. The presented results demonstrate the successful derivation of a desirable decomposition for the non-adiabatic terms, achieved through a linear combination utilizing Pauli operators. This attribute confers significant advantages to its practical implementation within quantum computing algorithms.
\end{abstract}


\newpage
\section{Introduction}
\setcounter{footnote}{0}
Quantum computing has emerged as a dynamic and vibrant domain of research within the scientific community, primarily driven by notable achievements and advancements in applied quantum machine learning \citep*{Jeremy2015,Biamonte2017,Dunjko2016,Hernani_2021}, quantum simulation \citep{Feynman_1982,Georgescu_2014}, and optimization of circuits and systems \citep{Tad_2000,Niroula_2022}. Optimization problems have garnered particular attention, given their pervasive presence in various domains, including medicine \citep{Prajapati_2023}, economics \citep{Tang_2022}, logistics \citep{Weinberg_2023}, and numerous others \citep{Ambainis_2017,Lloyd_2018,Dey_2016}. Classical approaches to solving these problems from an industrial perspective often face challenges in terms of efficiency and speed, thereby motivating the exploration of quantum computing as a promising alternative. The escalating interest in these methods can be attributed primarily to recent experimental advancements. This surge in interest is particularly noticeable from an industrial and commercial perspective. Consequently, there is growing anticipation that both conventional computers and quantum computing, in general, could yield significant advantages, eventually achieving a state known as ``quantum supremacy'' \citep{Arute_2019}. This potential advantage and progress have, in turn, spurred developments in various scientific domains, wherein contemporary quantum computers serve as proof of concept. However, it is essential to underscore that the applicability of these quantum algorithms warrants extensive research and investigation, particularly in the current state of quantum computing, which is commonly referred to as the Noisy Intermediate-Scale Quantum (NISQ) era \citep{NISQ} whose defining characteristic is the utilization of quantum processors with capacities of up to 1000 qubits.

Hybrid classical-quantum algorithms leverage NISQ devices while offloading a portion of their computational workload onto classical devices, offering considerable potential for practical applications in the field of quantum computing. A prominent example worth highlighting is the Variational Quantum Eigensolver (VQE) \citep{Peruzzo_2014}. The primary objective of VQE is to determine the lowest energy quantum state through hybrid optimization, utilizing a designated Hamiltonian operator in conjunction with variational quantum circuits. The realm of quantum machine learning also falls under the purview of these algorithms, seeking to adapt classical algorithms to their quantum counterparts to enhance and expedite computations by capitalizing on the principles of quantum superposition, entanglement, and interference. Within this domain, one can identify supervised classification algorithms like binary classification and those based on Grover's search algorithm \citep{Glover_1996}. Notably, Grover's algorithm has demonstrated quadratic acceleration in solving problems such as $k$-medians \citep{Aimeur_2013} or $k$-nearest neighbors \citep{Nathan_2014,Feldman_2023}. On the other hand, an alternative methodology that has significantly progressed in the literature of this field and has laid the foundation for numerous studies is the Quantum Approximate Optimization Algorithm (QAOA) \citep{Farhi_2022,Digitized-counterdiabatic}. This approach presents a valuable alternative for tackling combinatorial optimization problems using shallow quantum circuits through classical optimization of the associated parameters. In recent literature, substantial endeavors have been dedicated to employing these methodologies to solve the ground-state challenges of physical systems \citep{Pagano_2020,Lyu_2023,Torta_2023}, reflecting the ongoing efforts to enhance and adapt these techniques for broader applications in quantum optimization.

In recent years, significant attention and interest have been directed towards the development of adiabatic quantum optimization (AQO) methodologies for confronting optimization problems \citep{Santoro_2006,Albash_2018} with direct practical implementations in the branches of physics and chemistry \citep{Farhi_2001,Bapst_2013,Liu_2018}. These algorithms begin by initializing a quantum system in the ground state of a known Hamiltonian. The system's Hamiltonian is then slowly evolved into one that represents the problem to be solved, with its ground state encoding the solution. Leveraging the adiabatic theorem \citep{Born1928,Avron_1988}, it becomes feasible to ensure that the quantum system remains in its instantaneous state of lowest energy, provided the evolution of the Hamiltonian is carried out in a sufficiently slow and gradual manner and within a sufficiently extended period of time. Nevertheless, implementing slow adiabatic evolution at the experimental level is typically not feasible, necessitating the development of methodologies that accelerate these processes. In pursuit of this objective, recent scientific literature puts forth various approaches based on Shortcuts To Adiabaticity (STA) \citep{Odelin_2019,Takahashi_2019}. These methodologies encompass diverse techniques, including fast-forward methods \citep{masuda2010fastforward,Setiawan_2019}, invariant-based inverse engineering \citep{Chen_2011,Song_2016}, and counterdiabatic (CD) protocols \citep{Demirplak_2005,Hatomura_2017}. Despite the noticeable progress in the first two methodologies, this study primarily centers around the CD protocols. These techniques are specifically designed to speed up the adiabatic evolution process from an initial Hamiltonian to a final Hamiltonian. This is achieved by incorporating non-adiabatic terms following Equation (\ref{eqintro:1}), which effectively nullify the transition amplitudes between any energy eigenstate of the original Hamiltonian \citep{Berry}. Consequently, the quantum system undergoes an accelerated adiabatic evolution in practical applications. The resulting Hamiltonian by including the CD term is given by

\begin{equation}
\bm{\mathcal{H}}(t):=\bm{\mathcal{H}}_{\text{AD}}(t)+\bm{\mathcal{H}}_{\text{CD}}(t).
\label{eqintro:1}
\end{equation}

The operator designated as $\bm{\mathcal{H}}_{\text{AD}}(t)$ in (\ref{eqintro:1}) will be tailored to facilitate the preparation of its ground energy state at the initial time of the evolution. Nevertheless, the main challenge of the CD protocols lies in the accurate and plausible determination of the operator encompassing the non-adiabatic terms of the process, denoted here as $\bm{\mathcal{H}}_{\text{CD}}(t)$. In general, the computation and acquisition of this operator are exceedingly intricate tasks, particularly when dealing with many-body quantum systems \citep{portfolio_optimization}. As a customary approach in related literature, a time-dependent external parameterization, denoted as $\bm{\lambda}(t)$, is introduced to which the operators \citep{GeneratingSTA} are dependent (explained in detail in Section \ref{subsec:phys_back}). Efforts have been directed towards a method for the approximate determination of this operator, leading to significant progress, such as the development of the Nested Commutator (NC) methodology \citep{floquet}. This advancement has led to recent contributions, exemplified by \citep{Digitized-counterdiabatic,Digitized-counterdiabatic_2}. Within this framework, the computation of these terms is simplified into a series of commutators involving $\bm{\mathcal{H}}_{\text{AD}}(t)$ and its derivatives concerning the mentioned external parameterization $\bm{\lambda}(t)$. As a result, the non-adiabatic terms of these protocols are obtained approximately through an expansion in orders, where the complexity of obtaining them rises accordingly with the number of particles in the system and the order considered in the expansion. Nevertheless, these methodologies may exhibit problem-dependent characteristics, as their escalating complexity in non-trivial physical scenarios might necessitate the adoption of alternative perspectives to approach the issue at hand.

In a distinct domain, within the rapid progression of the computer science field, the realm of Deep Learning (DL) has achieved prominence as a highly potent instrument for constructing diverse models, owing to its direct applicability in domains such as image processing \citep{Andina_2018}, natural language generation and processing \citep{NLP}, time series prediction and classification \citep{timeseries}, among a plethora of other possibilities. Present-day technologies encompass Recurrent Neural Networks (RNNs) \citep{RNNs}, Long Short-Term Memory architectures (LSTMs) \citep{LSTMs}, the well-known transformers \citep{transformers}, sparse and submanifolds convolutional networks \citep{sparseconv} utilized for images with limited information load, among other advanced techniques. The Physics-Informed Neural Networks (PINNs) methodology has emerged as a highly intriguing DL application within the realm of physical sciences since its first appearances in the literature \citep{seminal_pinns_2,seminal_pinns}. This approach aims to address specific problems by employing neural networks as powerful universal approximators for systems of Partial Differential Equations (PDEs) that govern the physics describing the evolution. Due to their remarkable potential as numerical solvers, PINNs have garnered considerable attention and established themselves as a viable alternative to conventional numerical solving algorithms. Extensive efforts have been undertaken in diverse branches of physics to apply this methodology and its adaptations. These fields encompass classical hydrodynamics \citep{Zhiping_2020}, relativistic hydrodynamics \citep{GA_PINNs}, electrodynamics \citep{Jianxin_2022}, chemistry \citep{Ji_2021}, and many others. PINNs demonstrate their utility wherever differential equations are employed to describe the underlying physics of a given scenario. Their ability to unravel complex physical phenomena and offer numerical solutions has positioned them as promising tools for tackling intricate problems across various scientific domains. Consequently, the motivation to employ this methodology for addressing the challenge of CD protocols in quantum systems arises organically. The investigation of potential applications of PINNs in quantum systems and circuits becomes a natural course of study.

In this paper, we present an innovative approach for designing the counterdiabatic terms in CD protocols. Our proposed method lays on the utilization of PINNs, thereby representing a direct application of DL. We introduce a PINN-based methodology without supplementary alterations, enabling it to effectively resolve the underlying physics of the problem. Through the neural network, we obtain both the counterdiabatic terms and the temporal parameterization $\lambda(t)$, as expounded in Section \ref{subsec:phys_back}. Additionally, we directly explore the experimental feasibility by decomposing the non-adiabatic operator into tensor products of the set of Pauli and identity matrices. This approach offers a comprehensive and direct means to address the experimental applicability of the method.

The rest of the paper is structured as follows: Section \ref{sec:back} provides an introduction to the foundational theoretical framework concerning the operation of baseline PINNs. It further presents a comprehensive exposition of the theoretical framework under consideration, encompassing CD protocols and the specific problem that serves as the motivation for our research. Additionally, a thorough literature review of prior work in this domain is presented. Section \ref{sec:methods} delves into a meticulous presentation of the adapted PINN methodology, particularized to address our specific case, while taking into account all pertinent physical factors that the network must conform to. In Section \ref{sec:results}, we present notable outcomes obtained through our methodology and juxtapose them with previous findings in the field. Furthermore, we conduct comparisons and explore the scalability of the proposed methodology. Finally, Section \ref{sec:discuss} serves as a concluding segment, summarizing the principal conclusions drawn from our research and offering insights into potential avenues for future investigations.

\section{General Concepts}
\label{sec:back}
\subsection{Physics-Informed Neural Networks}
\label{subsec:PINN}
The fundamental approach employed in PINNs methodologies \citep{seminal_pinns} involves leveraging neural networks as powerful tools for approximating functions and solving physical problems by fitting sets of differential equations, known as PDEs. PINNs derive their name from the fact that they incorporate physical knowledge through the incorporation of inductive biases. These biases are manifested in various aspects of the methodology, including the design of the underlying neural network architecture, the formulation of appropriate cost functions (losses), and other characteristics that aim to ensure or enhance the convergence of the neural model. The underlying algorithm of these networks leverages the automated differentiation capabilities found in contemporary frameworks \citep{auto_diff} to construct differential equations based on the output variables obtained from the network. These variables are essential for computing the specific problem at hand. By performing calculations, a minimization process is subsequently employed, guided by a designated loss function, to update the trainable parameters of the architecture. Consequently, this adjustment aligns the network with the requirements of the physical framework.

Taking a broad perspective while maintaining generality, let us denote by $\bm{\bm{\mathcal{U}}}:=\bm{\mathcal{U}}(t,\bm{x})$ a collection of physical variables that serve as the output of the neural network. These variables, along with their derivatives, are the components of a system of PDEs defined within a domain of interest $\Omega$ over a specific time interval $[0,T]$. Consequently, it is possible to write the following definition:

\begin{equation}
\mathcal{F}\left(t,\bm{x};\frac{\partial\bm{\mathcal{U}}}{\partial t},\frac{\partial^{2}\bm{\mathcal{U}}}{\partial t^{2}},\ldots;\frac{\partial\bm{\mathcal{U}}}{\partial x_{1}},\ldots,\frac{\partial\bm{\mathcal{U}}}{\partial x_{D}};\frac{\partial^{2}\bm{\mathcal{U}}}{\partial x_{1}\partial x_{1}},\ldots,\frac{\partial^{2}\bm{\mathcal{U}}}{\partial x_{1}\partial x_{D}},\ldots\right)=0,\quad\bm{x}=(x_{1},\ldots,x_{D})\in\Omega,\quad t\in[0,T],
\label{eq1}
\end{equation}

where $D$ corresponds to the spatial dimension of the problem and $\Omega\subset\mathbb{R}^{D}$. The operator $\mathcal{F}$ defined in Equation (\ref{eq1}) can be conceptualized as the comprehensive collection of physical constraints inherent to the system, which must be fulfilled in order to satisfy the underlying PDEs. It is worth noting that, in addition to these constraints, supplementary limitations can be established, such as the initial conditions that dictate the evolution of the system, or potential boundary conditions that may influence the behavior of the system at the spatial boundaries of the domain, namely,

\begin{equation}
\mathcal{IC}(t,\bm{x})=0,\qquad(t,\bm{x})\in\{0\}\times\Omega.
\label{eq2}
\end{equation}
\begin{equation}
\mathcal{B}(t,\bm{x})=0,\qquad(t,\bm{x})\in(0,T]\times\partial\Omega.
\label{eq3}
\end{equation}

In addition to the aforementioned conditions, other factors can be taken into consideration, such as imposing additional constraints on the final time (final conditions), or at specific points of significant physical significance within the spatiotemporal framework. Furthermore, if there are actual experimental measurements available for a subset of the domain, they can also be incorporated. Consequently, each of these physical conditions represents a segment of a priori knowledge regarding the physical scenario that can be integrated into the cost function as separate terms (referred to as ``soft enforcement''), as denoted with $\mathcal{L}_{i}$ in Equation (\ref{eq4}):

\begin{equation}
\mathcal{L}:=\omega_{\mathcal{F}}\mathcal{L}_{\mathcal{F}}+\sum_{i}\omega_{i}\mathcal{L}_{i},
\label{eq4}
\end{equation}

where $\mathcal{L}_{\mathcal{F}}$ corresponds to the metric pertaining to the underlying system of PDEs, while the collection $(\omega_{\mathcal{F}}, \omega_{i}, \ldots)$ represents the weights assigned to each term within the mixture. The neural architecture employed in this methodology yields a set of essential physical variables, denoted as $\bm{\mathcal{U}}(t,\bm{x};\Theta):=\bm{\mathcal{U}}_{\Theta}(t,\bm{x})$, where $\Theta$ encompasses all the trainable parameters of the network that are updated during the training process. Consequently, the output aims to closely align with the corresponding real-world values:

\begin{displaymath}
\bm{\mathcal{U}}_{\Theta}(t,\bm{x})\approx\bm{\mathcal{U}}(t,\bm{x}).
\end{displaymath}

The constraints expressed in (\ref{eq2}) and (\ref{eq3}) can be transformed into cost functions by employing a suitable difference measurement metric, such as \textit{Mean Squared Error} (MSE) or similar approaches. The determination of these cost functions together with $\mathcal{L}_{\mathcal{F}}$ can be outlined as follows.

\begin{equation}
\mathcal{L}_{\mathcal{IC}}:=\frac{1}{N_{\mathcal{IC}}}\sum_{\{0\}\times\Omega}\mathopen|\bm{\mathcal{U}}_{\Theta}(0,\bm{x})-\bm{\mathcal{U}}(0,\bm{x})\mathclose|^{2},
\label{eq5}
\end{equation}

\begin{equation}
\mathcal{L}_{\mathcal{B}}:=\frac{1}{N_{\mathcal{B}}}\sum_{(0,T]\times\partial\Omega}\mathopen|\bm{\mathcal{U}}_{\Theta}(t,\bm{x}\in\partial\Omega)-\bm{\mathcal{U}}(t,\bm{x}\in\partial\Omega)\mathclose|^{2},
\label{eq6}
\end{equation}

\begin{equation}
\mathcal{L}_{\mathcal{F}}:=\frac{1}{N_{\mathcal{F}}}\sum_{(0,T]\times\Omega}\underbrace{\sum_{k}\mathopen|\mathcal{R}_{k}(t,\bm{x};\Theta)\mathclose|^{2}}_{\mathopen|\mathcal{F}\mathclose|^{2}}.
\label{eq7}
\end{equation}

Here, the set $(N_{\mathcal{IC}},N_{\mathcal{B}},N_{\mathcal{F}})$ represents the number of points in the sample for each respective domain considered. Additionally, $\mathcal{R}_{k}$ denotes the fulfillment of specific physical constraints imposed at the level of the PDE. Through the utilization of the fundamental methodology employed in PINN models, it becomes feasible to smoothly enforce the imposed constraints by adjusting the PDEs associated with the given problem. Nonetheless, alternative approaches, such as the ``hard enforcement'' method \citep{hard_enforcement}, propose compelling the neural network to enforce predetermined constraints from the onset of training by modifying the output of the network. This technique necessitates incorporating several constraints, but it entails establishing a specific dependence on the problem at hand. Other researchers achieve a certain level of independence concerning the set of weights $(\omega_{\mathcal{R}},\omega_{i},...)$ in (\ref{eq4}) through the application of the Augmented Lagrangian method \citep{lag_mult}. This technique involves updating multipliers with corresponding names during training in accordance with the degree of violation for each respective condition.

\subsection{Physical background}
\label{subsec:phys_back}
Combinatorial and optimization problems are of great interest in many industry and research fields \citep{Digitized-counterdiabatic,Digitized-counterdiabatic_2}. Adiabatic Quantum Computing (AQC) algorithms are used to solve this kind of problems and they are expected to outperform classical computers in the current NISQ era \citep{OptimizingCD,sels2017minimizing}. In this paradigm, we can prepare a Hamiltonian in an initial ground state, driving or mixing Hamiltonian, and evolve it towards a desired final operator, whose ground state encodes the solution to the underlying problem, or it can also be the solution itself \citep{Digitized-counterdiabatic_2,portfolio_optimization}. This initial or \text{driving} Hamiltonian operator should be easy to prepare and evolve. We shall commence by establishing the Hamiltonian operator associated with the adiabatic transition of the system, $\bm{\mathcal{H}}_{\text{AD}}(t)$, which is characterized by its energy eigenvalues, $E_{n}(t)$, and corresponding eigenstates, $\ket{n(t)}$,  as determined by:

\begin{equation}
\bm{\mathcal{H}}_{\text{AD}}(t)\ket{n(t)}=E_{n}(t)\ket{n(t)}.
\label{eq:H_AD_acting}
\end{equation}

A time-dependent Hamiltonian, as the one defined in (\ref{eq:H_AD_acting}), generally could lead to modifications in the quantum states it governs. Nevertheless, when these changes are sufficiently minute, analytically tractable, and under controlled conditions, the adiabatic theorem \citep{Born1928} ensures that the system, assuming non-degeneracy in energy, will preserve its proximity to the initial energy state throughout its temporal evolution. Considering the aforementioned perspective, it is consistently feasible to formulate a Hamiltonian operator, denoted as $\bm{\mathcal{H}}(t)$, that exhibits a direct correlation with $\bm{\mathcal{H}}_{\text{AD}}(t)$ and accurately reproduces the temporal progression of its eigenstates. In other words, it possesses transition amplitudes between energy levels that are precisely zero \citep{Berry}. This operator will be designed to satisfy the following Schrödinger equation:

\begin{displaymath}
i\hbar\:\partial_{t}\ket{\psi_{n}(t)}=\bm{\mathcal{H}}(t)\ket{\psi_{n}(t)},
\end{displaymath}

where $\ket{\psi_{n}(t)}$ can be defined in terms of the corresponding eigenstate of the operator in (\ref{eq:H_AD_acting}), $\ket{n(t)}$. These states represent each one of the $n$ states driven by $\bm{\mathcal{H}}_{\text{AD}}(t)$ in a certain particular system. Through rigorous derivation and computation, one can obtain a plausible analytical representation for the operator $\bm{\mathcal{H}}(t)$. For a detailed deducement, the interested reader is encouraged to refer to the works of \citep{Berry,Kato,Nenciu}. This operator, defined in (\ref{eq:H_kets_definition}), effectively governs the evolution of energy states within the framework of $\bm{\mathcal{H}}_{\text{AD}}(t)$, ensuring the absence of transitions between them.

\begin{equation}
\bm{\mathcal{H}}(t)=\underbrace{\sum_{n}\ket{n}E_{n}\bra{n}}_{\bm{\mathcal{H}_{\text{AD}}}}+\underbrace{i\hbar\:\sum_{n}\left(\ket{\partial_{t}n}\bra{n}-\braket{n|\partial_{t}n}\ket{n}\bra{n}\right)}_{\bm{\mathcal{H}_{\text{CD}}}}:=\bm{\mathcal{H}}_{\text{AD}}(t)+\bm{\mathcal{H}}_{\text{CD}}(t).
\label{eq:H_kets_definition}
\end{equation}

The Hamiltonian operator $\bm{\mathcal{H}}_{\text{CD}}(t)$ corresponding to the second term in (\ref{eq:H_kets_definition}), can be interpreted as the operator responsible for capturing the counterdiabatic effects during system evolution. These effects facilitate the acceleration of the underlying system dynamics by introducing additional accessible degrees of freedom allowing to reach the same results as the entire adiabatic evolution of the system and eliminating any possible transition between eigenstates \citep{sels2017minimizing,COLD}, as previously stated. These frameworks, which are the well-known as counterdiabatic Protocols, effectively expedite the adiabatic processes by introducing novel terms that precisely offset any excitations that may arise during system acceleration. These have been recently developed as part of the STA methods. With this in mind, it is feasible to extend the theoretical deduction by incorporating a set of time-dependent external parameters, denoted as $\bm{\lambda}(t)$, upon which all our operators will be dependent \citep{GeneratingSTA}. By doing so, when retrieving Equation (\ref{eq:H_kets_definition}), it follows that the temporal derivatives of the $\ket{n}$ states can be written as follows,

\begin{equation}
\ket{\partial_{t}n}=\frac{d\bm{\lambda}}{dt}\ket{\bm{\nabla}_{\bm{\lambda}}n}.
\label{eq:derivative_n}
\end{equation}

Equation (\ref{eq:derivative_n}) allows us to redefine the operator $\bm{\mathcal{H}}(t)$ as

\begin{equation}
\bm{\mathcal{H}}(t):=\bm{\mathcal{H}}_{\text{AD}}(t)+\underbrace{\frac{d\bm{\lambda}}{dt}\bm{\mathcal{A}}_{\text{CD}}(t)}_{\bm{\mathcal{H}}_{\text{CD}}},\qquad\bm{\mathcal{A}}_{\text{CD}}(t):=i\hbar\:\sum_{n}\left(\ket{\bm{\nabla}_{\lambda}n}\bra{n}-\braket{n|\bm{\nabla}_{\lambda}n}\ket{n}\bra{n}\right).\footnote{While the operators in this formulation exhibit time dependence through $\lambda$, for the sake of notation simplicity, we will denote the time dependency explicitly.}
\label{eq:H_definition}
\end{equation}

In general, this parameterization could encompasse a collection of values $\bm{\lambda}:=(\lambda_{1},...,\lambda_{N})$. However, to align with the latest literature in the field, we will confine ourselves to a single scalar function. Consequently, $\lambda(t)$, which usually receives the name of \textit{scheduling function} in the literature, carries out the counterdiabatic driving by terms of its temporal derivative. Indeed, it is evident that in the limit of zero velocity, $\left|\frac{d\lambda}{dt}\right|\rightarrow 0$, the Hamiltonian in Equation (\ref{eq:H_definition}) simplifies to the adiabatic operator, as anticipated during its construction. When extrapolating this understanding to contemporary digitized versions of algorithms within the realm of quantum circuits \citep{portfolio_optimization}, it is customary to initiate the process with a Hamiltonian which presents a ground state of energy that can be readily prepared in practical settings, denoted as $\bm{\mathcal{H}}_{\text{initial}}$. Under adiabatic conditions, it is imperative for this operator to undergo a gradual transformation over time, eventually converging to the target Hamiltonian corresponding to the specific problem under investigation, written as $\bm{\mathcal{H}}_{\text{problem}}$.

\begin{equation}
\bm{\mathcal{H}}_{\text{AD}}(t):=\left(1-\lambda(t)\right)\bm{\mathcal{H}}_{\text{initial}}+\lambda(t)\:\bm{\mathcal{H}}_{\text{problem}}.
\label{eq:Had}
\end{equation}

In Equation (\ref{eq:Had}), the scheduling function can once again be identified. Within the time interval $t\in[t_{\text{min}},t_{\text{max}}]$, it is necessary for $\lambda(t)$ to fulfill the conditions $\lambda(t_{\text{min}})=0$ and $\lambda(t_{\text{max}})=1$ based on its functional definition, which enables the interpolation of the process from the initial Hamiltonian to the desired endpoint of the process.

On the other hand, the $\bm{\mathcal{A}}_{\text{CD}}(t)$ operator (also written in the literature as $\bm{\mathcal{A}}_{\lambda}$) is defined as the Adiabatic Gauge Potential. Obviously, this operator should fulfill the hermiticity condition, i.e., $\bm{\mathcal{A}}_{\text{CD}}=\bm{\mathcal{A}}_{\text{CD}}^\dag$, in order to be interpreted as a physical observable. It is also easy to show that the potential $\bm{\mathcal{A}}_{\text{CD}}(t)$ satisfies the condition (\ref{eq:Comm}), which is equivalent to minimizing the action $\mathcal{S}$ defined in Equation (\ref{eq:S}) for the Hilbert-Schmidt operator $\bm{\mathcal{G}}_{\lambda}$ \citep{GeneratingSTA,Geometry}. Consequently, Equation (\ref{eq:Comm}) can be understood as the Euler-Lagrange equation resulting from the minimization of the physical action.

\begin{equation}
\mathcal{S}=\text{Tr}\left[\bm{\mathcal{G}}_{\lambda}^{2}\right],\qquad\bm{\mathcal{G}}_{\lambda}\left(\bm{\mathcal{A}}_{\text{CD}}\right)=\partial_{\lambda}\bm{\mathcal{H}}_{\text{AD}}+\frac{i}{\hbar}\left[\bm{\mathcal{A}}_{\text{CD}},\bm{\mathcal{H}}_{\text{AD}}\right].
\label{eq:S}
\end{equation}

\begin{equation}
\left[i\hbar\frac{\partial\bm{\mathcal{H}}_{\text{AD}}}{\partial\lambda} - [\bm{\mathcal{A}}_{\text{CD}}, \bm{\mathcal{H}}_{\text{AD}}], \bm{\mathcal{H}}_{\text{AD}}\right]=0.\footnote{To provide clarity and consistency, from now on we will exclusively employ Planck units, wherein the reduced Planck constant, denoted as $\hbar$, is established as unity (i.e., $\hbar=1$).}
\label{eq:Comm}
\end{equation}

This term should establish a connection between the aforementioned external parameter $\lambda(t)$ and instantaneous eigenstates. Finding this exact CD terms is not easy without spectral information of the system and they are usually approximated using the NC approach \citep{Digitized-counterdiabatic,Digitized-counterdiabatic_2,portfolio_optimization} or through Variational Circuits \citep{OptimizingCD}. This lead to a set of possible local CD terms and different techniques that have been developed to determine which of them are the most adequate for the particular problem. Nonetheless, these methods even though they are of great interest and may constitute the state-of-the-art approaches, they are still approximations. Consequently, there could be other relevant terms and aspects that are not considered within these methodologies.

\subsection{Quantum circuit design for the $\mathrm{H_{2}}$ ground state problem}
\label{subsec:quantum_circuit}

The main numerical application of this paper will be to find the ground state of the $\mathrm{H_{2}}$ molecule in the STO-3G basis assuming different bond distances where, in particular, this STO-3G basis corresponds to a minimal set that uses three Gaussian functions to approximate the atomic orbitals \citep{Szabo-Ostlund}. This can be described with the 2-qubit Full Configuration Interaction (FCI) Hamiltonian in Equation (\ref{eq:Hfci}), where the value of the coefficients vary with the bond distances. This Hamiltonian has been obtained using the well-know \textit{Qiskit} module \citep{Qiskit}. In a different domain, it is well-known that the Pauli matrices comprise a set of Hermitian and unitary matrices, namely $\{\sigma_{\text{X}},\sigma_{\text{Y}},\sigma_{\text{Z}}\}\in\mathcal{M}_{2\times 2}(\mathbb{C})$, which are highly suitable for representing quantum operators and physical observables. These matrices, together with the identity matrix $\bm{\mathcal{I}}$ (also called $\sigma_{0}$), form an orthogonal basis of the Hilbert space of $2\times 2$ Hermitian matrices. Similarly, when dealing with a system of $N_{Q}$ qubits, the matrix space of Hermitian $\mathcal{M}_{2^{N_{Q}}\times 2^{N_{Q}}}(\mathbb{C})$ matrices can be decomposed by means of tensor products involving the aforementioned set (the interested reader is referred to \citep{yanofsky2008quantum} Chapter 2 and \citep{rieffel2011quantum} Chapter 3). This procedure, referred to as the Pauli basis expansion, enables us to achieve a comprehensive representation of any Hermitian operator on a system comprising a certain amount of qubits. From the perspective of quantum circuits, this approach offers a structured and concise depiction of quantum operations and physical observables, thereby facilitating the analysis and manipulation of quantum systems.

\begin{equation}
\bm{\mathcal{H}}_{\text{FCI}}=c_{0}\bm{\mathcal{I}}^{(1)}\otimes\bm{\mathcal{I}}^{(2)}+c_{1}\:\bm{\mathcal{I}}^{(1)}\otimes\sigma_{\text{Z}}^{(2)}+c_{2}\:\sigma_{\text{Z}}^{(1)}\otimes\bm{\mathcal{I}}^{(2)}+c_{3}\:\sigma_{\text{Z}}^{(1)}\otimes\sigma_{\text{Z}}^{(2)}+c_{4}\:\sigma_{\text{X}}^{(1)}\otimes\sigma_{\text{X}}^{(2)}+c_{5}\:\bm{\mathcal{I}}^{(1)}\otimes\bm{\mathcal{I}}^{(2)}.
\label{eq:Hfci}
\end{equation}

This corresponds to $\mathcal{H}_{\text{problem}}$ in Equation (\ref{eq:Had}). In this context, the numeric superscripts enclosed in parentheses mean that the written operator pertain to the specific qubit under consideration thereby resulting in distinct matrices associated with each one. The symbol $\otimes$ denotes the Kronecker product. Furthermore, the first and last coefficients are written separately with the same operator since they correspond to different atoms in the molecule, even though they have the same interaction. The FCI Hamiltonian in Equation (\ref{eq:Hfci}) can also be written in matrix form Equation (\ref{eq:Hfcimatrix}).

\begin{equation}
\bm{\mathcal{H}}_{\text{FCI}}=
\begin{bmatrix}
c_{0}+c_{1}+c_{2}+c_{3}+c_{5} & 0 & 0 & c_{4} \\
0 & c_{0}-c_{1}+c_{2}-c_{3}+c_{5} & c_{4} & 0  \\
0 & c_{4} & c_{0}+c_{1}-c_{2}-c_{3}+c_{5} & 0  \\
c_{4} & 0 & 0 & c_{0}-c_{1}-c_{2}+c_{3}+c_{5}
\end{bmatrix}.
\label{eq:Hfcimatrix}
\end{equation}

As detailed before, we start from a \textit{driving} Hamiltonian that should be easy to prepare. For this particular problem, the Hartee-Fock (HF) approximation Hamiltonian (\ref{eq:HHF}) is used, with its matrix form written in Equation (\ref{eq:HHFmatrix}). It is easy to see that both the FCI and the HF Hamiltonian are real-valued since they are composed only with the identity matrix, $\bm{\mathcal{I}}$, and the Pauli matrices $\sigma_{\text{X}}$ and $\sigma_{\text{Z}}$.

\begin{equation}
\bm{\mathcal{H}}_{\text{HF}}=p_{0}\:\bm{\mathcal{I}}^{(1)}\otimes\bm{\mathcal{I}}^{(2)}+p_{1}\:\bm{\mathcal{I}}^{(1)}\otimes\sigma_{\text{Z}}^{(2)}+p_{2}\:\sigma_{\text{Z}}^{(1)}\otimes\bm{\mathcal{I}}^{(2)}+p_{3}\:\sigma_{\text{Z}}^{(1)}\otimes\sigma_{\text{Z}}^{(2)}+p_{4}\:\bm{\mathcal{I}}^{(1)}\otimes\bm{\mathcal{I}}^{(2)}.
\label{eq:HHF}
\end{equation}

\begin{equation}
\bm{\mathcal{H}}_{\text{HF}}=\begin{bmatrix}
p_{0}+p_{1}+p_{2}+p_{3}+p_{4} & 0 & 0 & 0 \\
0 & p_{0}-p_{1}+p_{2}-p_{3}+p_{4} &  0 & 0  \\
0 & 0 & p_{0}+p_{1}-p_{2}-p_{3}+p_{4} & 0  \\
0 & 0 & 0 & p_{0}-p_{1}-p_{2}+p_{3}+p_{4}
\end{bmatrix}.
\label{eq:HHFmatrix}
\end{equation}

The HF method approximates the ground state of the system using a single Slater determinant. The contributions of both the initial and final Hamiltonian operators in the adiabatic evolution can be determined using advanced numerical algorithms \citep{Qiskit}. As stated before, the primary objective is to solve the adiabatic temporal evolution while considering non-adiabatic terms. The goal is to accelerate this evolution without allowing unexpected transitions between energy states. By incorporating the adiabatic operator defined in Equation (\ref{eq:Had}) into the total Hamiltonian operator described in Equation (\ref{eq:H_definition}), and considering that the initial and final Hamiltonians are represented as shown in Equations (\ref{eq:Hfci}-\ref{eq:HHFmatrix}) with $\bm{\mathcal{H}}_{\text{initial}}:=\bm{\mathcal{H}}_{\text{HF}}$ and $\bm{\mathcal{H}}_{\text{final}}:=\bm{\mathcal{H}}_{\text{FCI}}=\bm{\mathcal{H}}_{\text{problem}}$, we can establish the following PDE,

\begin{equation}
\bm{\mathcal{H}}(t):=\left(1-\lambda(t)\right)\bm{\mathcal{H}}_{\text{HF}}+\lambda(t)\bm{\mathcal{H}}_{\text{problem}}+\frac{d\lambda}{dt}\bm{\mathcal{A}}_{\text{CD}}(t).
\label{eq:H_PDE}
\end{equation}

\subsection{Related work}
\label{subsec:rel_work}
The primary objective of CD driving processes is to attain a gauge potential, $\bm{\mathcal{A}}_{\text{CD}}(t)$, by expediting the adiabatic process through a temporal velocity parameter, specifically the time derivative of the scheduling function, $\frac{d\lambda}{dt}$. The aforementioned operator possesses the property that the product $\frac{d\lambda}{dt}\bm{\mathcal{A}}_{\text{CD}}$ in Equation (\ref{eq:H_PDE}) fully mitigates the occurrence of non-adiabatic transitions that would otherwise manifest in a specific eigenstate $\ket{n(t)}$ under the total $\bm{\mathcal{H}}(t)$, which has undergone evolution from the initial state $\ket{n(t=0)}$ of the control Hamiltonian, $\bm{\mathcal{H}}_{\text{AD}}(t)$ \citep{Deffner_2014}. Therefore, the aim of the related methodologies is to ensure precise compensation of the adiabatic terms in the evolutionary dynamics within the moving frame \citep{Demirplak_2005}. This potential, comprising the non-adiabatic compensations, represents an operator that is typically intricate to derive. While an exact numerical construction is possible for certain systems with a small number of particles, resolving it for systems with numerous bodies (a substantial number of qubits) becomes exceedingly challenging, if not impossible, due to the requirement of diagonalizing the resulting Hamiltonian observable across the entire Hilbert space. Furthermore, $\bm{\mathcal{A}}_{\text{CD}}$ often entails the emergence of highly intricate and non-local couplings, thus limiting its implementation to specific cases \citep{Zwick_2014,Takahashi_2016,Diao_2018}. Part of the research focused on a robust obtaining of the potential $\bm{\mathcal{A}}_{\text{CD}}$ has led to works such as \citep{Masuda_2008} where fast-forward methods are presented through which it is possible to obtain an objective wave function in a shorter time. This investigation is conducted within the field of microscopic quantum mechanics, aiming to delve into the macroscopic domain primarily governed by the Schrödinger equation. These inquiries are expounded upon in references such as \citep{masuda2010fastforward, Torrontegui_2012}. Despite these notable progressions, there presently exists no viable approach to extend these studies to the context of many-body systems. Consequently, these methodologies are not applicable to complex problem sets. On the other hand, several recent studies such as \citep{OptimizingCD} have suggested alternative approaches for addressing the aforementioned issues. The authors suggest to employ Variational Circuits (VC) in conjunction with classical optimizers to optimize the choice of CD terms, with these being treated as trainable parameters, being tested on the specific case of an Ising model of interactions with nearby neighbors and carrying out a comparison with some of the state-of-the-art QAOA methodologies \citep{Digitized-counterdiabatic,Wurtz_2022,Chai_2022}. Some assumptions are made regarding the form of the function $\lambda(t)$ thereby predefining a temporal parameterization, which may exhibit certain dependencies on the problem under investigation.

Returning to the latest approaches to derive $\bm{\mathcal{A}}_{\text{CD}}(t)$ as described in the existing literature, based on its definition in Equation (\ref{eq:H_definition}) and assuming that the parameterization is completely determined by the function $\lambda(t)$, by differentiation of Equation (\ref{eq:H_AD_acting}) \citep{Berry}, it is straightforward to arrive at

\begin{equation}
\bra{m}\bm{\mathcal{A}}_{\text{CD}}\ket{n}=i\braket{m|\partial_{\lambda}n}=-i\frac{\bra{m}\partial_{\lambda}\bm{\mathcal{H}}_{\text{AD}}\ket{n}}{E_{m}-E_{n}}.
\label{eq:mAn}
\end{equation}

However, it can be readily observed from Equation (\ref{eq:mAn}) that determining the operator $\bm{\mathcal{A}}_{\text{CD}}(t)$ can become highly intricate and computationally expensive, especially when dealing with many-body problems. This potential necessitates exact diagonalization of the operator $\bm{\mathcal{H}}_{\text{AD}}(t)$ over time and, additionally, the difference in energies $E_{m}-E_{n}$ can introduce complications, leading to divergent and mathematically ill-defined matrix elements. In an attempt to address these challenges, the authors in \citep{floquet} suggest employing the so-called method of NC for approximating the representation of $\bm{\mathcal{A}}_{\text{CD}}(t)$.

\begin{equation}
\bm{\mathcal{A}}_{\text{CD}}^{(l)}=i\sum_{k=1}^{l}\alpha_{k}\underbrace{[\bm{\mathcal{H}}_{\text{AD}},[\bm{\mathcal{H}}_{\text{AD}},\ldots[\bm{\mathcal{H}}_{\text{AD}},\partial_{\lambda}}_{2k-1}\bm{\mathcal{H}}_{\text{AD}}]]].
\label{eq:NC}
\end{equation}

Equation (\ref{eq:NC}) presents a methodology for obtaining an approximate numerical approach for the operator $\bm{\mathcal{A}}_{\text{CD}}$. In this expression, $l$ represents the order of the expansion, and the set of coefficients $\{\alpha_{k}\}_{k=1}^{l}$ can be determined by minimizing the action in Equation (\ref{eq:S}) specifically tailored for the $l$-th order. For further details and a comprehensive demonstration, interested readers are referred to the reference, where it is demonstrated that the exact potential $\bm{\mathcal{A}}_{\text{CD}}$ is recovered in the limit as $l$ tends towards infinity. Notwithstanding its utilization as an approximation limited to an expansion order $l$, this ansatz has proven to be a methodological framework for attaining cutting-edge outcomes in the field of research \citep{Digitized-counterdiabatic,Digitized-counterdiabatic_2,portfolio_optimization,DAQF,DCQAPF}. In the absence of an analytical solution to the problem of the CD protocol, the most compelling alternative at our disposal is to employ techniques rooted in DL, which possess the capacity to comprehend and acquire knowledge of the fundamental physics governing the problem.

\section{Methodology}
\label{sec:methods}
\subsection{PINN framework}
Our approach involves employing a Physics-Informed Neural Network (PINN) that incorporates a neural network structure comprising multiple fully connected dense layers, with the total number of layers denoted as $K$. This includes both the input and output layers. Each layer consists of a variable number of neurons, represented by $N_{k}$, and is characterized by a common output activation function denoted as $\sigma_{k}$. This activation function remains consistent across all neurons within a given layer after it is specified. Let $\bm{W}^{(k)}\in\mathbb{R}^{N_{k}\times N_{k-1}}$ be the weight matrix connecting the $(k-1)$-th layer to the $k$-th layer, and $\bm{b}^{(k)}\in\mathbb{R}^{N_{k}}$ be the bias vector for the $k$-th layer. Consequently, the output of the $k$-th layer, denoted as $\bm{\mathcal{U}}_{\Theta}^{(k)}$, can be expressed as the application of the activation function $\sigma_{k}$ to the weighted sum of the output of the previous layer, followed by the addition of the biases, as denoted in Equation (\ref{eq8}):

\begin{equation}
\bm{\mathcal{U}}_{\Theta}^{(k)}=\sigma_{k}\left(\bm{W}^{(k)}\bm{\mathcal{U}}_{\Theta}^{(k-1)}+\bm{b}^{(k)}\right)
\label{eq8}
\end{equation}

In this manner, contingent upon the specific problem under consideration, additional network parameters beyond weights and biases may be taken into account. Nevertheless, if only these two sets are considered, the trainable variables would be limited to those typically found in a conventional network, denoted as $\Theta:=\{\bm{W}^{(k)},\bm{b}^{(k)}\}_{1\leq k\leq K}$. Regarding the activation functions, particularly those employed at the output layer, they need to be tailored to suit the requirements of the specific physical problem at hand. For instance, certain physical variables may exhibit limitations within a defined range of values. An example of such a case is the scheduling function as described in (\ref{eq:H_PDE}), which is constrained to the interval $[0,1]$ by definition, or the velocity of a specific fluid within relativistic scenarios which is subject to an upper limit that cannot surpass the speed of light \citep{GA_PINNs}. In the context of our specific problem, the sole independent variable is time, thereby allowing us to exclude spatial dimensions as inputs to the neural model. Consequently, by considering (\ref{eq8}) and the time interval $t\in[0,T]$, we can express the ensemble of output variables of the architecture in the following manner:

\begin{equation}
\bm{\mathcal{U}}(t)\approx\bm{\mathcal{U}}_{\Theta}(t)=\sigma_{K}\left(\bm{\mathcal{U}}_{\Theta}^{(K)}\circ\sigma_{K-1}\circ\bm{\mathcal{U}}_{\Theta}^{(K-1)}\circ\ldots\circ\sigma_{1}\circ\bm{\mathcal{U}}_{\Theta}^{(1)}\right)(t)
\label{eq9}
\end{equation}

Equation (\ref{eq9}) showcases the mathematical representation of the approximation proposed by the underlying neural network in our methodology. This approach aims to closely resemble the actual physical solution following the completion of the training process. In this context, the symbol $\circ$ represents the composition operator. Recent research in the field asserts that PINNs enhance their performance by incorporating dynamic activation functions that vary with the training process and are distinct for each neuron \citep{adaptive_functions}. However, our focus lies primarily on establishing the definition of physical inductive biases within the network. Consequently, each layer $k$ will be associated with an activation function denoted as $\sigma_{k}$, which uniformly affects the output tensor of that particular layer.

\subsection{Inductive biases and optimization}
\label{subsec:optimization}

Our approach involves employing a methodology based on PINNs, which allows us to incorporate strong inductive biases and a priori knowledge into the neural network. This incorporation is intended to ensure that the underlying physics governing the problem is adequately satisfied. To achieve this objective, it is essential to consider that the output of the underlying network in our methodology will consist of a set of variables denoted as

\begin{equation}
\bm{\mathcal{U}}_{\Theta}(t):=\left(\lambda,\bm{\mathcal{A}}_{\text{CD}},\bm{\mathcal{C}}\right)_{\Theta}\footnote{Henceforth, the utilization of the symbol ``$\Theta$'' to denote the network prediction will be omitted and presumed understood, except in cases where the terminology may potentially result in misconceptions.}.
\label{eq12}
\end{equation}

Here, $\lambda\in\mathbb{R}$ denotes the scheduling function, while $\bm{\mathcal{A}}_{\text{CD}}\in\mathcal{M}_{2^{N_{Q}}\times 2^{N_{Q}}}(\mathbb{C})$ represents the counterdiabatic terms of the evolution, and $\bm{\mathcal{C}}\in\mathbb{R}^{4^{N_{Q}}}$ stands for the set of coefficients in which these counterdiabatic terms can be linearly decomposed attending to all the possible tensors that come from the Kronecker products of the possible combinations according to both the number of qubits considered and the set of identity and Pauli matrices, $\{\bm{I},\sigma_{\text{X}},\sigma_{\text{Y}},\sigma_{\text{Z}}\}$ (see \citep{nielsen2010quantum} Chapter 2.1). In general, $\bm{\mathcal{A}}_{\text{CD}}$ is an operator composed of complex terms and will be of size $2^{N_{Q}}\times 2^{N_{Q}}$ with $N_{Q}$ being the number of qubits into consideration. Moreover, notwithstanding that the dependency may not always be explicitly stated, all variables emerging from the PINN are contingent upon the input time. Hence, the network yields solutions for each of the considered inference times.

The underlying neural network will be optimized using the so-called ``soft enforcement'' technique, as introduced in equations (\ref{eq5},\ref{eq6},\ref{eq7}). We will adapt this methodology to our specific scenario of Hamiltonian dynamics, where the global cost function can be decomposed into multiple terms. Specifically, we will address the initial and final time conditions for an input time interval $t\in[t_{\text{min}},t_{\text{max}}]$ as described in Equations (\ref{eq10}) and (\ref{eq11}), respectively.

\begin{equation}
\mathcal{L}_{\mathcal{IC}}:=\frac{\omega_{\mathcal{IC},1}}{N_{\mathcal{IC}}}\sum_{\{t_{\text{min}}\}}\left|\lambda(t_{\text{min}})\right|^{2}+\frac{\omega_{\mathcal{IC},2}}{N_{\mathcal{IC}}}\sum_{\{t_{\text{min}}\}}\left|\bm{\mathcal{H}}(t_{\text{min}})-\bm{\mathcal{H}}_{\text{HF}}\right|^{2},
\label{eq10}
\end{equation}

\begin{equation}
\mathcal{L}_{\mathcal{FC}}:=\frac{\omega_{\mathcal{FC},1}}{N_{\mathcal{FC}}}\sum_{\{t_{\text{max}}\}}\left|\lambda(t_{\text{max}})-1\right|^{2}+\frac{\omega_{\mathcal{FC},2}}{N_{\mathcal{FC}}}\sum_{\{t_{\text{max}}\}}\left|\bm{\mathcal{H}}(t_{\text{max}})-\bm{\mathcal{H}}_{\text{problem}}\right|^{2}.
\label{eq11}
\end{equation}

In the definitions above, $(\omega_{\mathcal{IC},1},\omega_{\mathcal{IC},2})$ and $(\omega_{\mathcal{FC},1},\omega_{\mathcal{FC},2})$ are the weights of the mixture in the calculation of the total loss function and whose values will depend on the knowledge applied as well as on the problem being treated, while $N_{\mathcal{IC}}$ and $N_{\mathcal{FC}}$ represent the number of sample points at the initial and final instants, respectively. Regarding the scheduling function of the problem, denoted as $\lambda(t)$, this function shall delineate the progression of physical states subsequent to the introduction of counterdiabatic terms as stipulated in Equation (\ref{eq:H_PDE}). At the initial time instant, we shall impose a condition where $\lambda(t_{\text{min}})$ equals zero, i.e., $\lambda(t_{\text{min}})=0$. Similarly, at the terminal time instant, we would prescribe that $\lambda(t_{\text{max}})$ assumes a value of one, i.e., $\lambda(t_{\text{max}})=1$, as per its formal definition \citep{Digitized-counterdiabatic_2}. The aforementioned conditions outlined correspond to the physical limitations that are inherently necessary to be satisfied within our specific scenario. At the initial moment, denoted as $t=t_{\text{min}}$, we enforce the scheduling function to be zero, while ensuring that the resulting Hamiltonian operator (\ref{eq:H_PDE}) is equivalent to the one obtained through the Hartree-Fock method (see \citep{szabo1996modern} Chapter 2.2), defined as $\bm{\mathcal{H}}(t_{\text{min}})=\bm{\mathcal{H}}_{\text{HF}}$. Additionally, by incorporating counterdiabatic terms, our intention is to accelerate the adiabatic transition and reduce the computational complexity of the underlying circuits \citep{portfolio_optimization,Digitized-counterdiabatic}. This, however, does not impede our knowledge of the final Hamiltonian operator, denoted as $\bm{\mathcal{H}}_{\text{problem}}$, which can be computed via advanced numerical methods in the chemistry field \citep{Qiskit}. These conditions, combined with the requirement that the scheduling function must be equal to one at the conclusion of the evolution, collectively constitute the final conditions mandated by the methodology.

After establishing the initial and final inductive biases explicitly, it becomes crucial to elucidate the physics governing the intermediate time periods within the interval. Upon acquiring all the aforementioned components, we possess the necessary elements to construct the complete Hamiltonian operator for the given problem, as delineated in Equation (\ref{eq:H_PDE}). However, for the purpose of coherence, we reiterate the expression here to avoid any disruption in the logical progression.

\begin{equation}
\bm{\mathcal{H}}(t)=\bm{\mathcal{H}}_{\text{AD}}(t)+\frac{d\lambda}{dt}\bm{\mathcal{A}}_{\text{CD}}(t),\qquad\text{with}\qquad\bm{\mathcal{H}}_{\text{AD}}(t):=\left(1-\lambda(t)\right)\bm{\mathcal{H}}_{\text{HF}}+\lambda(t)\bm{\mathcal{H}}_{\text{problem}}.
\label{eq:H_t_definitive}
\end{equation}

As it is well-known, the Hamiltonian operators in their general form are operators comprising complex numbers, which necessitate adherence to the property of Hermiticity. By satisfying this requirement, the operators can be appropriately interpreted as physical observables, enabling extraction of values from energy states that are purely real. Consequently, it is obvious that our operators should possess Hermitian properties. Furthermore, it is imperative to impose the condition that the neural network yields the solution for the Gauge potential (counterdiabatic terms) that achieves the utmost reduction in the physical action within the given scenario. This entails selecting the solution that results in achieving

\begin{equation}
\frac{\delta\mathcal{S}(\bm{\mathcal{A}_{\text{CD}}})}{\delta\bm{\mathcal{A}}_{\text{CD}}}=0.
\label{eq:S_minimization}
\end{equation}

The minimization of the physical action represents a crucial requirement for ensuring that the employed methodology yields an optimal operator $\bm{\mathcal{A}}_{\text{CD}}(t)$ under specific conditions, encompassing local temporary development, robustness, availability, and experimental accessibility, among other factors. Research studies, such as \citep{sels2017minimizing} and related literature, demonstrate the impact of the action in recovering the Euler-Lagrange Equation (\ref{eq:Comm}). Consequently, demanding the neural network to minimize the action is entirely equivalent to defining the term associated with the physical loss, as described in (\ref{eq13}). Moreover, it is well-established that the temporal rate of change of the scheduling function, denoted as $\frac{d\lambda}{dt}:=\dot{\lambda}$, represents the velocity or rate at which the non-adiabatic components drive the evolution of the system in the presence of the total Hamiltonian operator. Consequently, when the derivative approaches zero, i.e., $\dot{\lambda}=0$, the conventional adiabatic Hamiltonian is recovered. However, it is undesirable for the counterdiabatic terms to greatly surpass the adiabatic counterparts, as their purpose is to expedite the process without exerting dominant influence. Hence, it is essential that the time derivative of $\lambda$ remains small, yet not entirely nullified. This particular information must be communicated to the PINN through Equation (\ref{eq14}).

\begin{equation}
\mathcal{L}_{\text{Least Action}}:=\frac{\omega_{\text{Action}}}{N_{\mathcal{F}}}\sum_{(t_{\text{min}},t_{\text{max}})}\left[i\frac{\partial\bm{\mathcal{H}}_{\text{AD}}(t)}{\partial\lambda}-\left[\bm{\mathcal{A}}_{\text{CD}}(t),\bm{\mathcal{H}}_{\text{AD}}(t)\right],\bm{\mathcal{H}}_{\text{AD}}(t)\right].
\label{eq13}
\end{equation}

\begin{equation}
\mathcal{L}_{\text{Adiabaticity}}:=\frac{\omega_{\text{Ad}}}{N_{\mathcal{F}}}\sum_{(t_{\text{min}},t_{\text{max}})}\left|\frac{d\lambda}{dt}\right|^{2}.
\label{eq14}
\end{equation}

As mentioned in Section \ref{subsec:quantum_circuit}, the set $\{\sigma_{0},\sigma_{\text{X}},\sigma_{\text{Y}},\sigma_{\text{Z}}\}\in\mathcal{M}_{2\times 2}(\mathbb{C})$ form an orthogonal basis of the Hilbert space of $2\times 2$ Hermitian matrices. Consequently, it is both logical and well-founded to seek the computation of the operator $\bm{\mathcal{A}}_{\text{CD}}$ as a composite of tensor products involving the previously introduced Pauli matrices for a certain system of qubits. By doing so, it is possible to construct a linear combination of tensor products involving these matrices, yielding a set of coefficients denoted as $\bm{\mathcal{C}}\in\mathbb{R}^{4^{N_{Q}}}$. Each element within this set represents the relative magnitude of each term within the operator. Therefore, the decomposition expressed in Equation (\ref{eq15}) would enable us to perform efficient simulations and facilitate a more accessible analysis of physical systems through the utilization of quantum circuit models.

\begin{equation}
\bm{\mathcal{A}}_{\text{CD}}^{'}(t):=\sum_{i,j,\ldots,N_{Q}\in\{0,\text{X},\text{Y},\text{Z}\}}\bm{\mathcal{C}}_{i,j,\ldots,N_{Q}}(t)\left(\sigma_{i}\otimes\sigma_{j}\otimes\ldots\otimes\sigma_{N_{Q}}\right).
\label{eq15}
\end{equation}

In this study, we employ $\bm{\mathcal{A}}_{\text{CD}}^{'}(t)$ to represent the non-adiabatic terms. This notation serves to distinguish this expansion, which takes the form of a linear combination, from the operator that serves as an output of the PINN. To achieve this objective, it is necessary to introduce an additional term into the loss function of the neural network, which directly affects the set of coefficients denoted as $\bm{\mathcal{C}}(t)$, as shown in  Equation (\ref{eq16}). Consequently, these terms are dynamically adjusted during the training process in order to construct the decomposition of the Gauge operator. By employing the specified procedure and adhering to the prescribed requirement, our methodology exhibits the capability to yield these scalar quantities not only at the initial and final moments but also throughout the entire temporal interval. This is attributable to the fact that these scalars, in a general sense, are functions contingent upon time.

\begin{equation}
\mathcal{L}_{\text{Coupling}}:=\frac{\omega_{\text{Coupling}}}{N_{\mathcal{F}}}\sum_{(t_{\text{min}},t_{\text{max}})}\left|\bm{\mathcal{A}}_{\text{CD}}(t)-\bm{\mathcal{A}}_{\text{CD}}^{'}(t)\right|^{2}.
\label{eq16}
\end{equation}

Once all the requisite components have been specified according to Equations (\ref{eq13}), (\ref{eq14}), and (\ref{eq15}), it becomes feasible to establish the loss term for our PINN, Equation (\ref{eq17}), which is solely linked to the underlying differential equations. In this context, the vector $(\omega_{\text{Action}},\omega_{\text{Adiabaticity}},\omega_{\text{Coupling}})$ denotes the weights employed in the combination process when constructing the resultant term. Thus, by considering Equation (\ref{eq4}) and acknowledging the pre-established mixture weights within our loss terms, the formulation of the final loss function, Equation (\ref{eq18}), becomes straightforward. This incorporates the loss associated with the PDEs outlined in Equation (\ref{eq17}), as well as the temporal constraints at the initial and final temporal steps, as indicated in (\ref{eq10}) and (\ref{eq11}). Consequently, the defined loss function serves as the physical metric that guides the optimization process, dictating the objective for the neural network within our methodology to minimize.

\begin{equation}
\mathcal{L}_{\mathcal{F}}:=\mathcal{L}_{\text{Least Action}}+\mathcal{L}_{\text{Adiabaticity}}+\mathcal{L}_{\text{Coupling}}.
\label{eq17}
\end{equation}

\begin{equation}
\mathcal{L}:=\mathcal{L}_{\mathcal{IC}}+\mathcal{L}_{\mathcal{FC}}+\mathcal{L}_{\mathcal{F}}.
\label{eq18}
\end{equation}

In addition, the network has not explicitly been required to satisfy the necessary condition of operators hermiticity even though it can be achieved by including an additional term in Equation (\ref{eq17}) that minimizes the difference $\left|\bm{\mathcal{A}}_{\text{CD}} - \bm{\mathcal{A}}_{\text{CD}}^{\dag}\right|^2$. Nonetheless, such a restriction is not obligatory and would be redundant for the PINN. If the coefficients $\bm{\mathcal{C}} \in \mathbb{R}^{N_Q}$ are defined as real, i.e., $\bm{\mathcal{C}} = \bm{\mathcal{C}}^{*}$, then it is evident from the decomposition of $\bm{\mathcal{A}}^{'}_{\text{CD}}$ in Equation (\ref{eq15}) that we recover $\bm{\mathcal{A}}^{'}_{\text{CD}} - \bm{\mathcal{A}}_{\text{CD}}^{'\dag}=0$ naturally, without necessitating any additional requirements. Therefore, the physical condition expressed in Equation (\ref{eq16}) is more than sufficient for the neural network to ensure the hermiticity of the operator $\bm{\mathcal{A}}_{\text{CD}}$, and since $\bm{\mathcal{H}}_{\text{HF}},\bm{\mathcal{H}}_{\text{problem}}\in\mathcal{M}_{2^{N_{Q}}\times 2^{N_{Q}}}(\mathbb{R})$ \citep{Digitized-counterdiabatic}, the hermiticity of the complete Hamiltonian operator is also ensured.

\begin{figure}[h]
\centering
\scalebox{0.85}{
\begin{tikzpicture}

\node[circle, draw=black, fill=white](t) at (0,0) {$t$};
\node[blue,text width=2cm](text1) at (-2.5,0.0) {Time domain as input};
\draw[->,black] (text1) -- (-0.5,0.0);

\node[circle, draw=black, fill=gray!20, minimum size=0.5cm](w1_1) at (1.5,1.25) {};
\node[circle, draw=black, fill=gray!20, minimum size=0.5cm](w1_2) at (1.5,0.25) {};
\node[minimum size=0.5cm, rotate=90, text width=0.25cm](w1_3) at (1.5,-0.50) {...};
\node[circle, draw=black, fill=gray!20, minimum size=0.5cm](w1_4) at (1.5,-1.25) {};

\node[circle, draw=black, fill=gray!20, minimum size=0.5cm](w2_1) at (2.5,1.25) {};
\node[circle, draw=black, fill=gray!20, minimum size=0.5cm](w2_2) at (2.5,0.25) {};
\node[minimum size=0.5cm, rotate=90, text width=0.25cm](w2_3) at (2.5,-0.50) {...};
\node[circle, draw=black, fill=gray!20, minimum size=0.5cm](w2_4) at (2.5,-1.25) {};

\node[minimum size=0.5cm, rotate=0, text width=0.25cm](w3_1) at (3.25,1.15) {...};
\node[minimum size=0.5cm, rotate=0, text width=0.25cm](w3_2) at (3.25,0.35) {...};
\node[minimum size=0.5cm, rotate=90, text width=0.25cm](w3_3) at (3.25,-0.5) {...};
\node[minimum size=0.5cm, rotate=0, text width=0.25cm](w3_4) at (3.25,-1.35) {...};

\node[circle, draw=black, fill=gray!20, minimum size=0.5cm](w4_1) at (4.0,1.25) {};
\node[circle, draw=black, fill=gray!20, minimum size=0.5cm](w4_2) at (4.0,0.25) {};
\node[minimum size=0.5cm, rotate=90, text width=0.25cm](w4_3) at (4.0,-0.50) {...};
\node[circle, draw=black, fill=gray!20, minimum size=0.5cm](w4_4) at (4.0,-1.25) {};

\node[circle, draw=black, fill=gray!20, minimum size=0.5cm](w5_1) at (5.0,1.25) {};
\node[circle, draw=black, fill=gray!20, minimum size=0.5cm](w5_2) at (5.0,0.25) {};
\node[minimum size=0.5cm, rotate=90, text width=0.25cm](w5_3) at (5.0,-0.50) {...};
\node[circle, draw=black, fill=gray!20, minimum size=0.5cm](w5_4) at (5.0,-1.25) {};

\draw[-,black] (t) -- (w1_1);
\draw[-,black] (t) -- (w1_2);
\draw[-,black] (t) -- (w1_4);

\draw[-,black] (w1_1) -- (w2_1);
\draw[-,black] (w1_1) -- (w2_2);
\draw[-,black] (w1_1) -- (w2_4);

\draw[-,black] (w1_2) -- (w2_1);
\draw[-,black] (w1_2) -- (w2_2);
\draw[-,black] (w1_2) -- (w2_4);

\draw[-,black] (w1_4) -- (w2_1);
\draw[-,black] (w1_4) -- (w2_2);
\draw[-,black] (w1_4) -- (w2_4);

\draw[-,dashed,black] (w2_1) -- (w4_1);
\draw[-,dashed,black] (w2_1) -- (w4_2);
\draw[-,dashed,black] (w2_1) -- (w4_4);

\draw[-,dashed,black] (w2_2) -- (w4_1);
\draw[-,dashed,black] (w2_2) -- (w4_2);
\draw[-,dashed,black] (w2_2) -- (w4_4);

\draw[-,dashed,black] (w2_4) -- (w4_1);
\draw[-,dashed,black] (w2_4) -- (w4_2);
\draw[-,dashed,black] (w2_4) -- (w4_4);

\draw[-,black] (w4_1) -- (w5_1);
\draw[-,black] (w4_1) -- (w5_2);
\draw[-,black] (w4_1) -- (w5_4);

\draw[-,black] (w4_2) -- (w5_1);
\draw[-,black] (w4_2) -- (w5_2);
\draw[-,black] (w4_2) -- (w5_4);

\draw[-,black] (w4_4) -- (w5_1);
\draw[-,black] (w4_4) -- (w5_2);
\draw[-,black] (w4_4) -- (w5_4);

\node[circle, draw=none, minimum size=0, inner sep=0] (invisibleDot1) at (6.25,0.75) {};
\node[circle, draw=none, minimum size=0, inner sep=0] (invisibleDot2) at (6.25,0.0) {};
\node[circle, draw=none, minimum size=0, inner sep=0] (invisibleDot_new) at (6.25,-0.75) {};

\node[draw=none, fill=none](Lambda) at (8.25,0.75) {$\lambda (t)$};
\node[draw=none, fill=none](A_CD) at (8.25,0.0) {$\bm{\mathcal{A}}_{\text{CD}} (t)$};
\node[draw=none, fill=none](C) at (8.25,-0.75) {$\bm{\mathcal{C}} (t)$};

\draw[->,black] (invisibleDot1) -- (Lambda);
\draw[->,black] (invisibleDot2) -- (A_CD);
\draw[->,black] (invisibleDot_new) -- (C);

\draw[-,black] (w5_1) -- (invisibleDot1);
\draw[-,black] (w5_2) -- (invisibleDot1);
\draw[-,black] (w5_4) -- (invisibleDot1);

\draw[-,black] (w5_1) -- (invisibleDot2);
\draw[-,black] (w5_2) -- (invisibleDot2);
\draw[-,black] (w5_4) -- (invisibleDot2);

\draw[-,black] (w5_1) -- (invisibleDot_new);
\draw[-,black] (w5_2) -- (invisibleDot_new);
\draw[-,black] (w5_4) -- (invisibleDot_new);

\node[blue,text width=1cm, font=\small](text2) at (6.75,1.75) {$Sigmoid(x)$};

\node [draw, minimum width=0.5cm, minimum height=0.5cm] (func1) at (7.0,1.25) {
\begin{tikzpicture}
\draw[scale=0.075,domain=-3:3,samples=100,smooth,variable=\x,blue] plot ({\x},{3/(1+exp(-3*\x))});
\end{tikzpicture}
};

\node[draw=none, fill=none](NoneAct) at (7.0,0.25) {$\varnothing$};

\node[draw=none, fill=none](NoneAct_2) at (7.0,-0.5) {$\varnothing$};

\draw [decorate, decoration = {brace}] (9.0,1.0) --  (9.0,-1.0) node [blue,midway,text width=1.5cm,xshift=1.25cm](text3)
{\footnotesize Physical outputs};

\node[circle, draw=none, minimum size=0, inner sep=0] (invisibleDot3) at (8.25,1.5) {};

\draw[-,black] (Lambda) -- (invisibleDot3);

\node[draw=none, fill=none](U) at (10.5,1.5) {$\bm{\mathcal{U}}_{\Theta}(t):=\left(\lambda,\bm{\mathcal{A}}_{\text{CD}},\bm{\mathcal{C}}\right)$};

\draw[->,black] (invisibleDot3) -- (U);

\node[circle, draw=none, minimum size=0, inner sep=0] (invisibleDot4) at (11.5,0.0) {};

\draw[-,black] (text3) -- (invisibleDot4);

\node[draw=black, fill=none](H) at (11.5,-2.0) {$\bm{\mathcal{H}}(t)=\underbrace{\left(1-\lambda(t)\right)\bm{\mathcal{H}}_{\text{HF}}+\lambda(t)\bm{\mathcal{H}}_{\text{problem}}}_{\bm{\mathcal{H}}_{\text{AD}}}+\frac{d\lambda}{dt}\bm{\mathcal{A}}_{\text{CD}}(t)$};

\draw[->,black] (invisibleDot4) -- (H);

\node[draw=none, fill=none](text4) at (12.0,-0.5) {PDE};

\node[circle, draw=none, minimum size=0, inner sep=0] (invisibleDot5) at (11.5,-4.0) {};

\draw[-,black] (H) -- (invisibleDot5);

\draw[thick, decorate, decoration={brace, amplitude=10pt}] (8.0,-3.0) -- (8.0,-5.0) coordinate[midway] (bracket);

\draw[-,black] (H) -- (invisibleDot5);

\draw[->,black] (invisibleDot5) -- (8.5,-4.0);

\node[draw=none, fill=none](L_IC) at (4.5,-3.25) {$\mathcal{L}_{\mathcal{IC}}$};

\node[draw=none, fill=none](L_FINAL) at (4.5,-4.0) {$\mathcal{L}_{\mathcal{FC}}$};

\node[draw=none, fill=none](L_R) at (4.5,-4.75) {$\mathcal{L}_{\mathcal{F}}=\mathcal{L}_{\text{Least Action}}+\mathcal{L}_{\text{Adiabaticity}}+\mathcal{L}_{\text{Coupling}}$};

\draw[thick, decorate, decoration={brace, amplitude=10pt, mirror}] (1.0,-3.0) -- (1.0,-5.0) coordinate[midway] (bracket);

\node[circle, draw=black, fill=none, font=\scriptsize](plus) at (0.25,-4.0) {$+$};

\node[diamond, draw=black, font=\scriptsize](text5) at (-1.5,-4.0) {$\mathcal{L}<\epsilon?$};

\draw[->,black] (plus) -- (text5);

\node[draw=none, fill=none](text6) at (-2.0,-4.75) {Yes};

\node[draw=none, fill=none](text6) at (-1.5,-5.5) {End};

\draw[->,black] (text5) -- (text6);

\node[draw=none, fill=none](text6) at (-2.0,-2.75) {No};

\node[circle, draw=none, minimum size=0, inner sep=0] (invisibleDot6) at (-1.5,-2.0) {};

\draw[-,black] (text5) -- (invisibleDot6);

\node[circle, draw=none, minimum size=0, inner sep=0] (invisibleDot7) at (0.0,-2.0) {};

\draw[-,black] (invisibleDot6) -- (invisibleDot7);

\draw[->,black] (invisibleDot7) -- (t);

\end{tikzpicture}
}
\caption{The methodology follows a general procedure where time, represented by the variable $t$, is the only independent physical variable considered, incorporated into our network as a tensor. The output of the methodology comprises three elements: the scalar variable of the scheduling function denoted as $\lambda(t)$; the operator representing the counterdiabatic terms of the process, expressed as $\bm{\mathcal{A}}_{\text{CD}}(t)$, with each of its components treated as an independent output; and the coefficients $\bm{\mathcal{C}}(t)$ denote the general decomposition coefficients of the operator $\bm{\mathcal{A}}_{\text{CD}}$ expressed as a linear combination of tensors resulting from all possible Kronecker products formed between the set of operators comprising the identity matrix and the Pauli operators. Subsequently, the computations required to construct the total Hamiltonian, denoted as $\bm{\mathcal{H}}(t)$, are carried out. Various physical constraints are imposed during these computations, including inductive biases. These constraints adhere to the principle of minimum action, satisfy the initial and final conditions, and ensure the hermiticity of the physical operators among other specifications. At each step of the training process, the total loss is calculated by aggregating the contributions from all imposed conditions, denoted as $\mathcal{L}$. This calculation continues until a specific training period is reached or until a predetermined error threshold is achieved.}
\label{Diagram}
\end{figure}
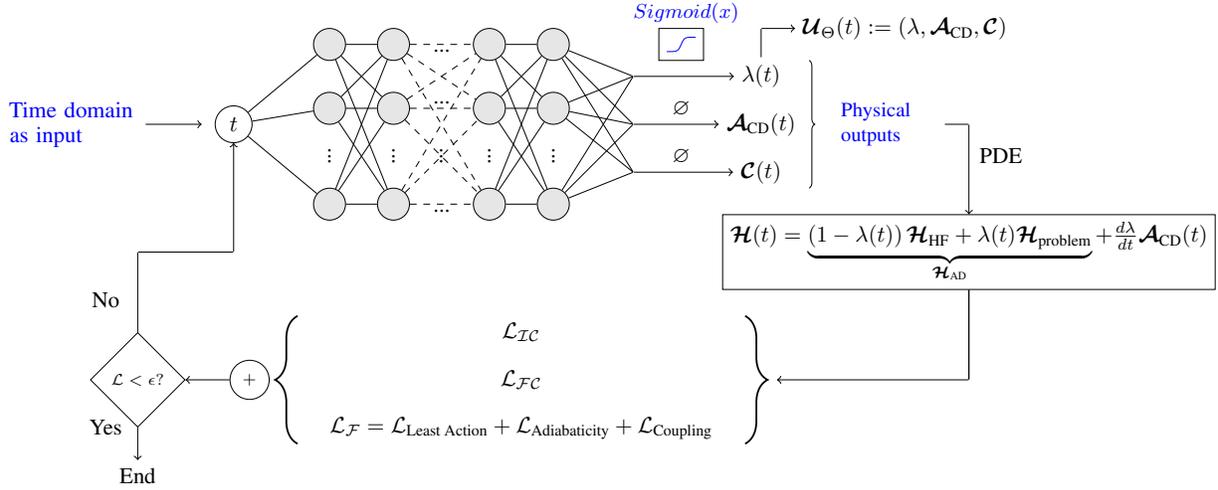

After considering all the relevant factors, a comprehensive description of our methodology can be provided. In order to accomplish this, we will utilize a visual aid in the form of a diagram (Figure \ref{Diagram}), and a step-by-step algorithm outlined in Algorithm \ref{algorithm1}. The initial step involves identifying the independent variable(s) that will influence our outcomes. In our particular case, we only have the temporal variable, denoted as $t$, defined within the interval $[t_{\text{min}},t_{\text{max}}]$, commonly specified as $[ t_{\text{min}},t_{\text{max}}]=[0,1]$. Consequently, we need to select an appropriate sampling method. Various methods are available, including uniform sampling with equidistant points, random sampling, Latin hypercube sampling \citep{LHS_1,LHS_2}, and Sobol sampling \citep{SobolSampling}. The choice of method is somewhat arbitrary, with greater impact when considering fewer time points, as its significance diminishes as the sample size approaches infinity. In particular, the Sobol sampling has exhibited significant advancements in prior inquiries within the relevant body of literature \citep{RAR_method}. In our investigation, we have employed this approach, which entails generating pseudo-random samples using powers of two. This technique results in a more homogeneous distribution of points within the space, with reduced overlap compared to completely random sampling. Nonetheless, it is important to acknowledge the inherent randomness involved in this approach. After generating the time domain, it serves as the input to the network, which will be constructed as a sequence of fully connected $K$ dense layers, encompassing both the input and output layers. The number of layers and the number of neurons in each layer ($N_{k}$) are hyperparameters that will be predetermined prior to training. While these parameters can be adjusted, it is anticipated that a relatively conventional configuration, such as 6 or 7 layers with 30 to 40 neurons each, will suffice for characterizing the counterdiabatic driving problem. Nevertheless, the complexity of individual problems may necessitate varying numbers of layers and/or neurons.

\begin{algorithm}[h]
\footnotesize
\KwIn{Physical domain: $t\in[t_{\text{min}},t_{\text{max}}]$.}
\KwOut{Set of variables produced by the PINN comprising the scheduling function, the operator controlling the counterdiabatic terms, and the coefficients of its associated Pauli decomposition, denoted as $\bm{\mathcal{U}}_{\Theta}(t):=(\lambda,\bm{\mathcal{A}}_{\text{CD}},\bm{\mathcal{C}})_{\Theta}$.}
\textbf{Specify the training domain:} Generate the temporal interval $t$ using \textit{Sobol sequence} \citep{SobolSampling}.\\
\textbf{Construct the underlying neural network} as a densely connected architecture consisting of $K$ layers, with a specific number of neurons per layer. Initialize the trainable parameters of the network, denoted as $\Theta=\{\bm{W}^{(k)},\bm{b}^{(k)}\}_{1\leq k\leq K}$, using the Glorot initialization method as described in the reference \citep{xavier_glorot}.\\
\textbf{By employing the principles of automatic differentiation}, compute the derivative $\frac{d\lambda}{dt}$ from the output and subsequently construct the operator $\bm{\mathcal{H}}(t)$ in accordance with the expression (\ref{eq:H_PDE}).\\
\textbf{Construct the initial and final losses in time} using the MSE metric as $\mathcal{L}_{\mathcal{IC}}$ (\ref{eq10}) and $\mathcal{L}_{\mathcal{FC}}$ (\ref{eq11}), respectively.\\
\textbf{Derive the comprehensive set of physical loss terms} associated with the differential equations, $\mathcal{L}$, incorporating considerations for the Principle of Least Action, adherence to adiabatic conditions, the decomposition in terms of Pauli tensor products and the hermiticity of the physical operators as a direct consequence. Subsequently, formulate the aggregate physical loss as follows:
\begin{displaymath}
\mathcal{L}:=\mathcal{L}_{\mathcal{IC}}+\mathcal{L}_{\mathcal{FC}}+\mathcal{L}_{\mathcal{F}}.
\end{displaymath}\\
\textbf{Update the set of network variables} by backpropagation, $\{\bm{W}^{(k)},\bm{b}^{(k)}\}_{1\leq k\leq K}$, that minimizes the total loss using a certain defined optimizer:
\begin{displaymath}
\Theta^{*}:=\text{arg min}\left(\mathcal{L}(\Theta)\right).
\end{displaymath}\\
\textbf{Repeat steps 3-6 for the required number of epochs} until the convergence is as desired or a maximum number of iterations is reached.\\
\caption{PINN algorithm for the counterdiabatic driving of a system of $N_{Q}$ qubits.}
\label{algorithm1}
\end{algorithm}

With the establishment of the input domain and the construction of the network, we are now capable of extracting the output and interpreting it as the tensor of physical variables denoted by $\bm{\mathcal{U}}_{\Theta}(t)=\left(\lambda,\bm{\mathcal{A}}_{\text{CD}},\bm{\mathcal{C}}\right)_{\Theta}$, where $\lambda\in\mathbb{R}$, $\bm{\mathcal{A}}_{\text{CD}}\in\mathcal{M}_{2^{N{Q}}\times 2^{N_{Q}}}(\mathbb{C})$, and $\bm{\mathcal{C}}\in\mathbb{R}^{4^{N_{Q}}}$. Subsequently, the derivative $\frac{d\lambda}{dt}$ can be straightforwardly computed using automatic differentiation, getting the Hamiltonian operator $\bm{\mathcal{H}}(t)$ as described in (\ref{eq:H_PDE}). Furthermore, the initial and final temporal constraints, defined in equations (\ref{eq10}) and (\ref{eq11}), are necessary. These constraints are accompanied by the calculation of the loss associated with the underlying PDEs stated in Equation (\ref{eq17}), which incorporates the physical constraints outlined in equations (\ref{eq13}), (\ref{eq14}), and (\ref{eq16}). These physical restrictions represent the fulfillment of the Principle of Least Action, the agreement with the adiabaticity, and the decomposition of the operator $\bm{\mathcal{A}}_{\text{CD}}(t)$ as previously explained. By combining all these components, the final loss metric $\mathcal{L}$ given in Equation (\ref{eq18}) is computed, and the set of trainable variables $\Theta$ is updated via backpropagation, minimizing the loss through an optimization process.

\section{Numerical experiments and results}
\label{sec:results}
In this section, various numerical results obtained through our DL-based methodology are presented, as outlined in Section \ref{sec:methods}. This approach has been applied to address the $\mathrm{H_{2}}$ molecule problem within the STO-3G basis, considering different bond distances between the particles and utilizing a 2-qubit representation. The initial and final Hamiltonian operators used for the evolution, as denoted by the PDE in (\ref{eq:H_PDE}), are listed in Table \ref{Table1} and were configured following the guidelines in \citep{Qiskit}. Since our numerical example involves a 2-qubit representation, these operators will possess a matrix size of $4\times 4$. Furthermore, it is pertinent to note that these operators are real-valued by definition, i.e., $\bm{\mathcal{H}}_{\text{HF}}, \bm{\mathcal{H}}_{\text{problem}}\in\mathcal{M}_{4\times 4}\left(\mathbb{R}\right)$. In a general context, and unless explicitly stated otherwise, all conducted trainings comprised a total of 500,000 epochs (iterations) dedicated to updating the $\Theta$ parameters of the used PINN. The training procedure utilized the \textit{PyTorch} module \citep{pytorch}, employing the \textit{Adam} optimizer \citep{adam_optimizer} with a fixed learning rate of $10^{-5}$. This learning rate was chosen to be sufficiently small, ensuring convergence without excessive numerical noise. Throughout all examples, the sampling method used has been the Sobol sampling \citep{SobolSampling} as stated in Section \ref{sec:methods}. The number of sampling points at the initial and final time instances of the evolution, denoted as $N_{\mathcal{IC}}$ and $N_{\mathcal{FC}}$ respectively, has remained constant at $2^{13}$. However, the number of points in the inner part of the interval, which is linked to the underlying PDE and represented as $N_{\mathcal{F}}$, was subject to variation and examination in the experiments. Moreover, if not otherwise specified, the neural architecture used for all the examples consists of six hidden layers with 30 neurons each.
\begin{table}[h]
\caption{Numerical configurations for the initial Hamiltonian operator, denoted as $\bm{\mathcal{H}}_{\text{HF}}$, and the final Hamiltonian operator, denoted as $\bm{\mathcal{H}}_{\text{problem}}$, obtained using the quantum computing library \textit{Qiskit} \citep{Qiskit}. These configurations correspond to the evolution of the molecule $\mathrm{H_{2}}$ in the STO-3G basis and are represented by their respective matrix descriptions given in Equations (\ref{eq:HHFmatrix}) and (\ref{eq:Hfcimatrix}), respectively.}
  \label{Table1}
  \centering
  \scriptsize
  \renewcommand{\arraystretch}{1.2}
  \begin{tabular}{ccccccccccc}
    \toprule
    & \multicolumn{4}{c}{$\bm{\mathcal{H}}_{\text{HF}}$} &  &                       & \multicolumn{4}{c}{$\bm{\mathcal{H}}_{\text{problem}}$} \\
    \cmidrule{2-5} \cmidrule{8-11}
    \multirow{2}{*}{$d=1.0\:\text{\AA}$} & -0.5490812 & 0           & 0          & 0           &  &                       & -0.5490812 & 0           & 0          & 0.19679058 \\
    & 0          & -1.0661087 & 0          & 0           &  &                       & 0          & -1.0661087 & 0.19679058 & 0          \\
    & 0          & 0          & 0.00400595 & 0           &  &                       & 0          & 0.19679058  & 0.00400595 & 0          \\
    & 0          & 0          & 0          & -0.5490812  &  &                       & 0.19679058 & 0           & 0          & -0.5490812 \\
    \midrule
    \multirow{2}{*}{$d=1.5\:\text{\AA}$} & -0.6610488 & 0           & 0          & 0           &  &                       & -0.6610488 & 0           & 0          & 0.22953594 \\
    & 0          & -0.91087353 & 0          & 0           &  &                       & 0          & -0.91087353 & 0.22953594 & 0          \\
    & 0          & 0           & -0.3944683 & 0           &  &                       & 0          & 0.22953594  & -0.3944683 & 0          \\
    & 0          & 0           & 0          & -0.6610488  &  &                       & 0.22953594 & 0           & 0          & -0.6610488 \\
    \midrule
    \multirow{2}{*}{$d=2.0\:\text{\AA}$} & -0.66539884 & 0           & 0          & 0           &  &                       & -0.66539884 & 0           & 0          & 0.25913846 \\
    & 0          & -0.7837927  & 0          & 0           &  &                       & 0          & -0.7837927  & 0.25913846 & 0          \\
    & 0          & 0           & -0.5412806 & 0           &  &                       & 0          & 0.25913846  & -0.5412806 & 0          \\
    & 0          & 0           & 0          & -0.66539884 &  &                       & 0.25913846 & 0           & 0          & -0.66539884 \\
    \midrule
    \multirow{2}{*}{$d=2.5\:\text{\AA}$} & -0.649429   & 0           & 0          & 0           &  &                       & -0.649429   & 0           & 0          & 0.28221005 \\
    & 0          & -0.7029436  & 0          & 0           &  &                       & 0          & -0.7029436  & 0.28221005 & 0          \\
    & 0          & 0           & -0.5944048 & 0           &  &                       & 0          & 0.28221005  & -0.5944048 & 0          \\
    & 0          & 0           & 0          & -0.649429   &  &                       & 0.28221005 & 0           & 0          & -0.649429   \\
    \bottomrule
  \end{tabular}
  \vspace{0.25cm}
  \end{table}

\begin{figure}[!htbp]
\centering
\includegraphics[scale=0.360]{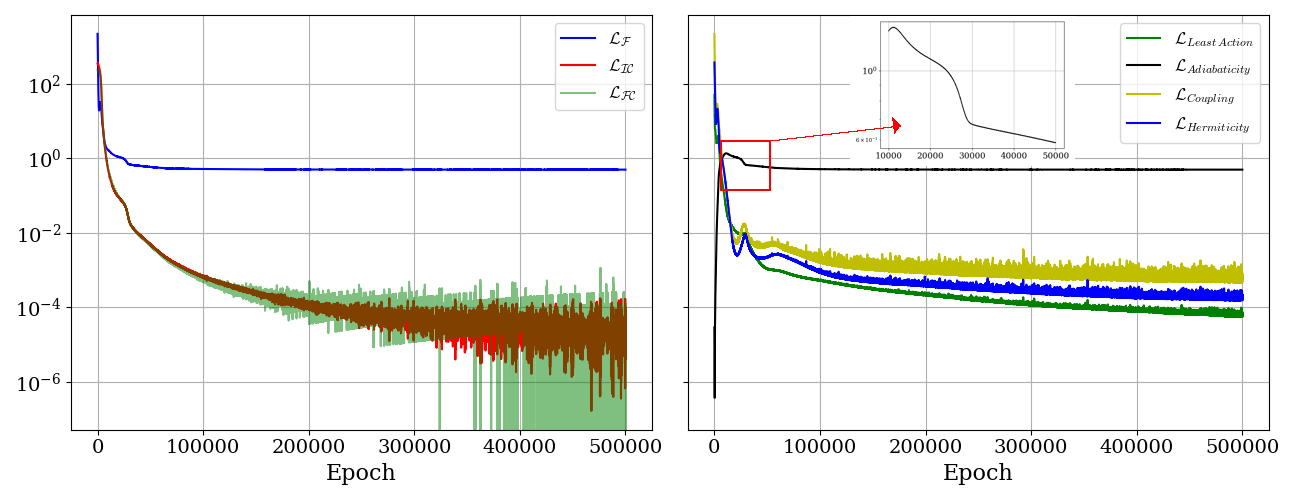} 
\caption{Analysis of the evolution of the loss function during the training process. On the left we illustrate the dynamic changes of the components contributing to the total loss, $\mathcal{L}$, as defined in Equation (\ref{eq18}). On the right side of the graph, each individual constituent of the loss, $\mathcal{L}_{\mathcal{F}}$, is presented, corresponding to the different physical aspects under consideration. It is important to note that the loss term $\mathcal{L}_{\text{Hermiticity}}$ is included in the plot, although it remains undefined and unused throughout the training. This term quantifies the discrepancy between $\bm{\mathcal{A}}_{\text{CD}}$ and its adjoint operator but is solely provided for visualization purposes, tracking its reduction relative to $\mathcal{L}_{\text{Coupling}}$ due to their mathematical equivalence.}
\label{fig:Losses}
\end{figure}

To illustrate initial general results, Figure \ref{fig:Losses} presents the physical loss functions employed for optimizing the neural network, as defined in Section \ref{sec:methods}. It is essential to emphasize that each training process would exhibit in general distinctive loss curves, influenced by factors such as the number of qubits used for representation and the specific molecular system being studied, including the considered bond distances between molecules, among other features. However, the figure showcases the cost function for the specific scenario under investigation: the $\mathrm{H_{2}}$ molecule, represented by a 2-qubit system in the STO-3G basis, with specifications outlined in Section \ref{subsec:quantum_circuit}. The left subfigure displays the three constituents comprising the total loss $\mathcal{L}$ (\ref{eq18}), namely, $\mathcal{L}_{\mathcal{F}}$, $\mathcal{L}_{\mathcal{IC}}$, and $\mathcal{L}_{\mathcal{FC}}$, corresponding to the residual conditions of the PDE, the initial conditions over time, and the final conditions, respectively. The latter two diminish rapidly, converging to magnitudes on the order of $10^{-4}$ or even lower, especially evident for $\mathcal{L}_{\mathcal{FC}}$, where the error is so minute that any variation induces substantial noise on a logarithmic scale. Conversely, the residual loss $\mathcal{L}_{\mathcal{F}}$ (\ref{eq17}) encompasses three internal terms, generating discernible tension among them, which impedes its reduction to such small orders of magnitude. This becomes more evident in the right subfigure, which provides a detailed breakdown of these terms. It is evident that the loss term responsible for minimizing the Principle of Least Action or, equivalently, the Euler-Lagrange equations (\ref{eq13}), attains the lowest value, thus ensuring a highly satisfactory minimization of the action and, consequently, the attainment of a potential gauge $\bm{\mathcal{A}}_{\text{CD}}(t)$ that adheres closely to its prescribed guidelines. Additionally, the graphical representation includes a loss term denoted as $\mathcal{L}_{\text{Hermiticity}}$, which assesses the quadratic discrepancy between $\bm{\mathcal{A}}_{\text{CD}}(t)$ and its corresponding operator, defined in a similar way to how the other terms are. However, this factor is not incorporated in the total loss; instead, it serves as a representation allowing us to confirm that minimizing the loss $\mathcal{L}_{\text{Coupling}}$ (\ref{eq16}) through the decomposition of the potential gauge directly enforces its hermiticity, given the strict real-valued nature of the coefficients $\bm{\mathcal{C}}(t)$ involved in the decomposition (\ref{eq15}). Thus, for all the examples shown in this article, the mix weights considered for each of the individual loss terms have been preset as:

\begin{equation}
\left(\omega_{\mathcal{IC}},\omega_{\mathcal{FC}},\omega_{\text{Action}},\omega_{\text{Ad}},\omega_{\text{Coupling}}\right)=\left(10^{3},10^{3},10^{2},5\times 10^{-1},2.5\times 10^{2}\right),
\label{eq:mixture_weights}
\end{equation}

where $\omega_{\mathcal{IC},1}=\omega_{\mathcal{IC},2}=\omega_{\mathcal{IC}}$ and $\omega_{\mathcal{FC},1}=\omega_{\mathcal{FC},2}=\omega_{\mathcal{FC}}$ in Equations (\ref{eq10}) and (\ref{eq11}). These weights play a crucial role in determining the relative emphasis that the neural network places on each term, essentially representing their respective priorities. For our specific objectives, ensuring prompt and robust satisfaction of the initial and final conditions in time is of crucial importance. Consequently, we set $\omega_{\mathcal{IC}}$ and $\omega_{\mathcal{FC}}$ to significantly larger values than $\omega_{\text{Action}}$, $\omega_{\text{Ad}}$, and $\omega_{\text{Coupling}}$. Following the same thoughts, the minimization of the action is also essential in our methodology as it leads us to identify the most appropriate operator $\bm{\mathcal{A}}_{\text{CD}}(t)$ for the system \citep{sels2017minimizing}, and we express it through its linear combination (\ref{eq15}). Consequently, these two weights will be considerably higher, yet an order of magnitude lower than the first two. Particularly, $\omega_{\text{Coupling}}$ is considered slightly higher, as this condition inherently encompasses the constraint on the hermiticity of the operator. Lastly, the condition of adiabaticity will be assigned the lowest weight, reflecting our intent to recover the limit of the adiabatic theory without permitting it to dominate the overall metric.

\FloatBarrier

We could continue our analysis of the right-hand subfigure by specifically putting our attention on the curve that corresponds to the loss of recovery of the adiabatic theory, denoted as $\mathcal{L}_{\text{Adiabaticity}}$ (\ref{eq14}) and represented in a solid black curve. This loss exhibits a marginal decline, commencing from a value proximal to zero at the onset of the training process. Subsequently, it rises to approximately $10^{2}$ before descending to around $10^{0}$ and eventually maintaining a constant level of magnitude for the duration of the evolution. The physical reason behind this phenomenon is clear: while the theory necessitates a decrease in the adiabatic speed, $\frac{d\lambda}{dt}$, the PINN fails to recognize the appropriateness of taking it to values lower than what is observed in our results. This is primarily attributed to the presence of multiple terms within the loss function, not solely related to adiabaticity recovery, thereby creating inherent tension in defining the physical problem as a whole. Consequently, the solution predicted by the methodology does not significantly benefit from further reductions in the adiabatic speed. The fast saturation of adiabatic recovery is the main cause of the considerably elevated value for $\mathcal{L}_{\mathcal{F}}$ observed in the left subfigure, underscoring the importance of isolating and analyzing each term separately. Thus, a zoom has been applied to $\mathcal{L}_{\text{Adiabaticity}}$ between 30,000 and 50,000 iterations, a range during which the $\lambda$ function transitions from a sigmoidal to a linear behavior. Consequently, it assumes a temporal derivative of 1, i.e., $\frac{d\lambda}{dt}=1$. A more comprehensive discussion on this topic will be presented in the following section. It is important to note that unless otherwise specified, all the results shown in this section have been carried out for the system formed by 2 qubits representing the $\mathrm{H_{2}}$ molecule in the STO-3G basis, following the guidelines of Section \ref{subsec:quantum_circuit}.

\subsection{Scheduling function, $\lambda(t)$}
The output of our methodology is triple. Firstly, it enables the retrieval of the scheduling function $\lambda$ as a time-dependent variable. Secondly, it facilitates the extraction of the matrix components $(i,j)$ of the potential gauge $\bm{\mathcal{A}}_{\text{CD}}(t)$, considering also the temporal variable, and it also enables the obtaining of the $\bm{\mathcal{C}}(t)$ components of its decomposition through time. This achievement is made possible by the utilization of a cost metric defined on the underlying PDE comprising three distinct terms, as stated in Equation (\ref{eq17}). Each of these terms compels the neural network to update one of the outputs while adhering to the imposed physical constraints. In particular, the term $\mathcal{L}_{\text{Adiabaticity}}$ plays a crucial role in enforcing the adherence of the function $\lambda(t)$ to the physical evolution, thereby recovering, in the most accurate approximation, the adiabatic theory that has been empirically demonstrated to hold true. Among all the terms taken into account within the total cost function, the neural network exhibits the lowest attention towards fulfilling the requirement of recovering the fully adiabatic theory, $\mathcal{L}_{\text{Adiabaticity}}$. As illustrated in Figure \ref{fig:Losses}, our PINN ceases to actively optimize this particular term after approximately 50,000 iterations. Consequently, the scheduling function $\lambda(t)$ converges to an optimal solution wherein its time derivative, representing the adiabatic velocity, maintains an approximately constant value of 1 throughout the temporal evolution. As a consequence, the optimal form of the $\lambda$ function predicted for the system is the one that adheres to the condition $\lambda(t)=t$ for the evolution, as governed by the counterdiabatic differential Equation (\ref{eq:H_PDE}).

\begin{figure}[!htbp]
\centering
\includegraphics[scale=0.60]{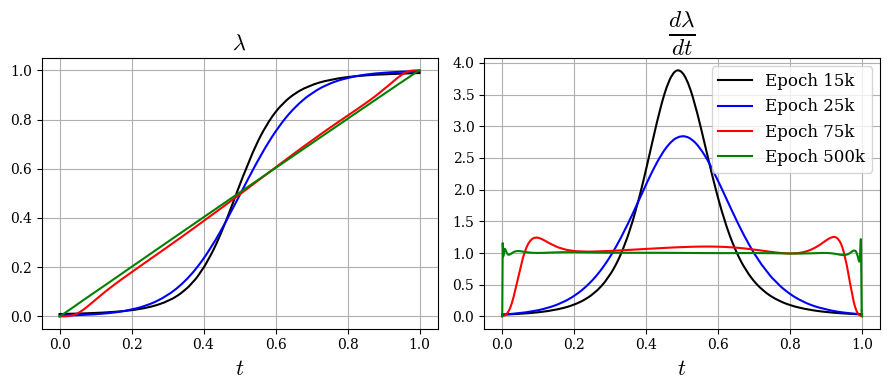} 
\caption{Analysis of the scheduling function $\lambda$ and its temporal derivative for distinct iterations (epochs) of the training process. Observations reveal that during the initial stages of neural optimization, $\lambda$ exhibits characteristics resembling a sigmoidal function. However, as the training advances, it converges towards a linear behavior $\left(\frac{d\lambda}{dt}\approx 1\right)$.}
\label{fig:Lambda}
\end{figure}

This phenomenon is exemplified in Figure \ref{fig:Lambda}, where we show the values of $\lambda(t)$ and its temporal derivative in the left and right subfigures, respectively, for different training steps in both cases concerning the 2-qubit system representing the $\mathrm{H_{2}}$ molecule. As observed, even after 15,000 and 25,000 epochs, the function maintains a sigmoidal shape, a widely employed representation in the existing literature \citep{Digitized-counterdiabatic,Digitized-counterdiabatic_2}. This sigmoidal form is considered appropriate from a theoretical perspective due to its derivative taking the form of a ``bell curve'', facilitating the adiabatic terms $\bm{\mathcal{A}}_{\text{CD}}$ to present higher values at intermediate values during time evolution while effectively being turned off at the initial and temporal instants. It should be noted, however, that this sigmoidal shape for $\lambda$ emerges predominantly during the early phases of the training thereby helping in the driving of the convergence of the methodology towards a more optimal solution. From a theoretical point of view, our ultimate goal is to recover the fully adiabatic theory while incorporating the non-zero presence of counterdiabatic terms. Consequently, our neural network converges towards a function $\lambda$ that precisely matches the temporal variable $t$. This outcome signifies the restoration of the original formulation of counterdiabatic driving, as explained in \citep{Berry}, thereby undoing the temporal parameterization of physical operators through a specific set of parameters $\bm{\lambda}(t)$, which, in our case, corresponds to a single scalar parameterization.

\FloatBarrier

\begin{figure}[!htbp]
\centering
\includegraphics[scale=0.60]{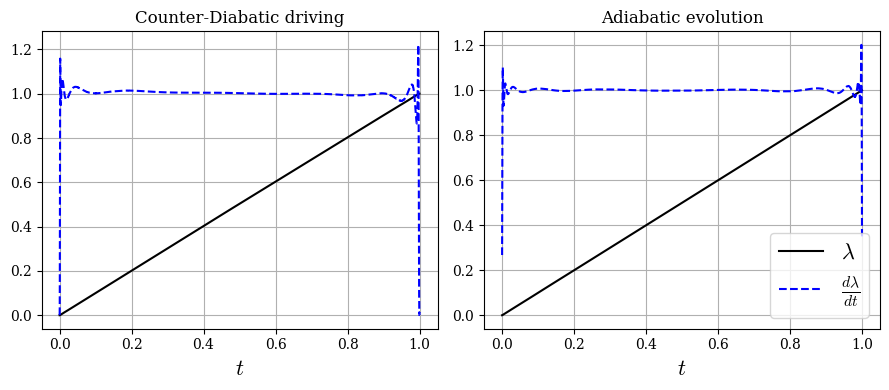} 
\caption{Evolution over time of the \textit{scheduling function} $\lambda(t)$ (in solid black) along with its derivative (in dashed blue) predicted by our methodology for the $\mathrm{H_{2}}$ molecule in the STO-3G basis set using the CD protocol on the left. On the right, the same is depicted for a fully adiabatic evolution according to expression (\ref{eq:Had}).}
\label{fig:Lambda2}
\end{figure}

Through this procedure, our objective is to begin with a differential equation containing non-zero counterdiabatic terms and then strive to recover the adiabatic theory in the limiting case. By doing so, we can obtain all the necessary results in accordance with the theory of adiabaticity. In Figure \ref{fig:Lambda2}, we present the profiles of $\lambda(t)$ and its time derivative for the counterdiabatic (CD) protocol from Equation (\ref{eq:H_PDE}) on the left, while on the right subfigure, we depict the same results but without considering the presence of counterdiabatic terms, i.e., working directly with the adiabatic theory (\ref{eq:Had}). It is evident that for both cases we obtain $\lambda(t)=t$, except for some numerical noise at the edges of the time interval, which is directly related to the nature of the automatic differentiation process \citep{auto_diff,cheney2007numerical,brown2012fourier}. Notably, during the initial stages of training, the neural network capitalizes on the sigmoid-shaped $\lambda(t)$, while simultaneously adjusting other physical conditions. This aids the network in achieving a more optimal solution in recovering the adiabatic theory.

\FloatBarrier

\subsection{Temporal evolution of the \texorpdfstring{$\bm{\mathcal{H}}(t)$}{H} operator and the energy levels}
\label{subsec:energy_levels}

\begin{figure}[!htbp]
\centering
\includegraphics[scale=0.44]{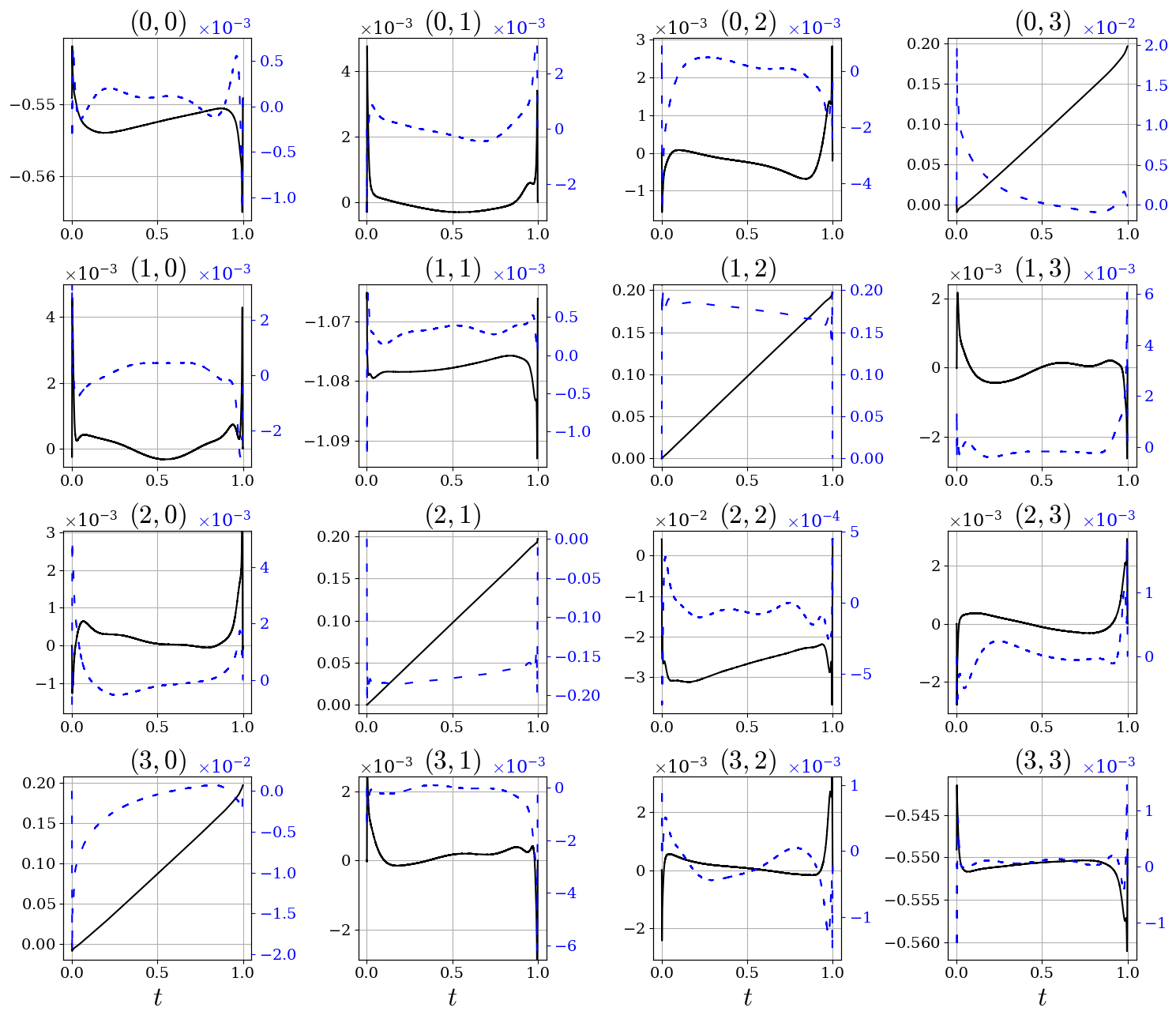} 
\caption{The evolution of the real and imaginary components of the Hamiltonian operator $\bm{\mathcal{H}}(t)$ for the $\mathrm{H_{2}}$ molecule is examined using a bond distance of $d=1.0\:\text{\AA}$. Black and blue colors have been used for real and imaginary parts, respectively, using dashed lines for the latter and a different scale for each one. The findings reveal substantial fluctuations in the values across various time scales. Notably, the natural symmetry and antisymmetry for the real and imaginary components, respectively, arise due to the hermiticity of the operator.}
\label{fig:Total_H}
\end{figure}

From the three outputs obtained from the network (see diagram in Figure \ref{Diagram}), all the necessary information can be derived. As an initial step, considering the potential gauge $\bm{\mathcal{A}}_{\text{CD}}(t)$ that minimizes the physical action of the system, the total Hamiltonian operator $\bm{\mathcal{H}}(t)$ can be computed. The time evolution of all its components is illustrated in Figure \ref{fig:Total_H}. Notably, $\bm{\mathcal{H}}\in\mathcal{M}_{2^{N_{Q}}\times 2^{N_{Q}}}\left(\mathbb{C}\right)$, where $N_{Q}$ represents the number of qubits, implying that for a 2-particle system, both $\bm{\mathcal{H}}(t)$ and the remaining operators will have a matrix size of $4\times 4$. In both depictions, the inherent hermiticity of the observable is evident; the real part of the components exhibits symmetry with respect to the diagonal, while the imaginary part displays complete antisymmetry, leading to diagonal elements being as close to zero as possible (around the order of $10^{-3}\sim 10^{-4}$). This condition can be further reinforced by increasing its relative weight, $\omega_{\text{Coupling}}$. In the mentioned figure, the real and imaginary parts of distinct components of the operator are depicted within the same graph, distinguished by black and blue colors, respectively, and with individual scales. By considering the fulfillment of hermiticity for $\bm{\mathcal{H}}(t)$, we can now extract its instantaneous eigenvalues, thereby obtaining information about the energy levels of the 2-qubit physical system representing the $\mathrm{H_{2}}$ molecule with which we are currently conducting our investigation.

\FloatBarrier

\begin{figure}[!htbp]
\centering
\includegraphics[scale=0.43]{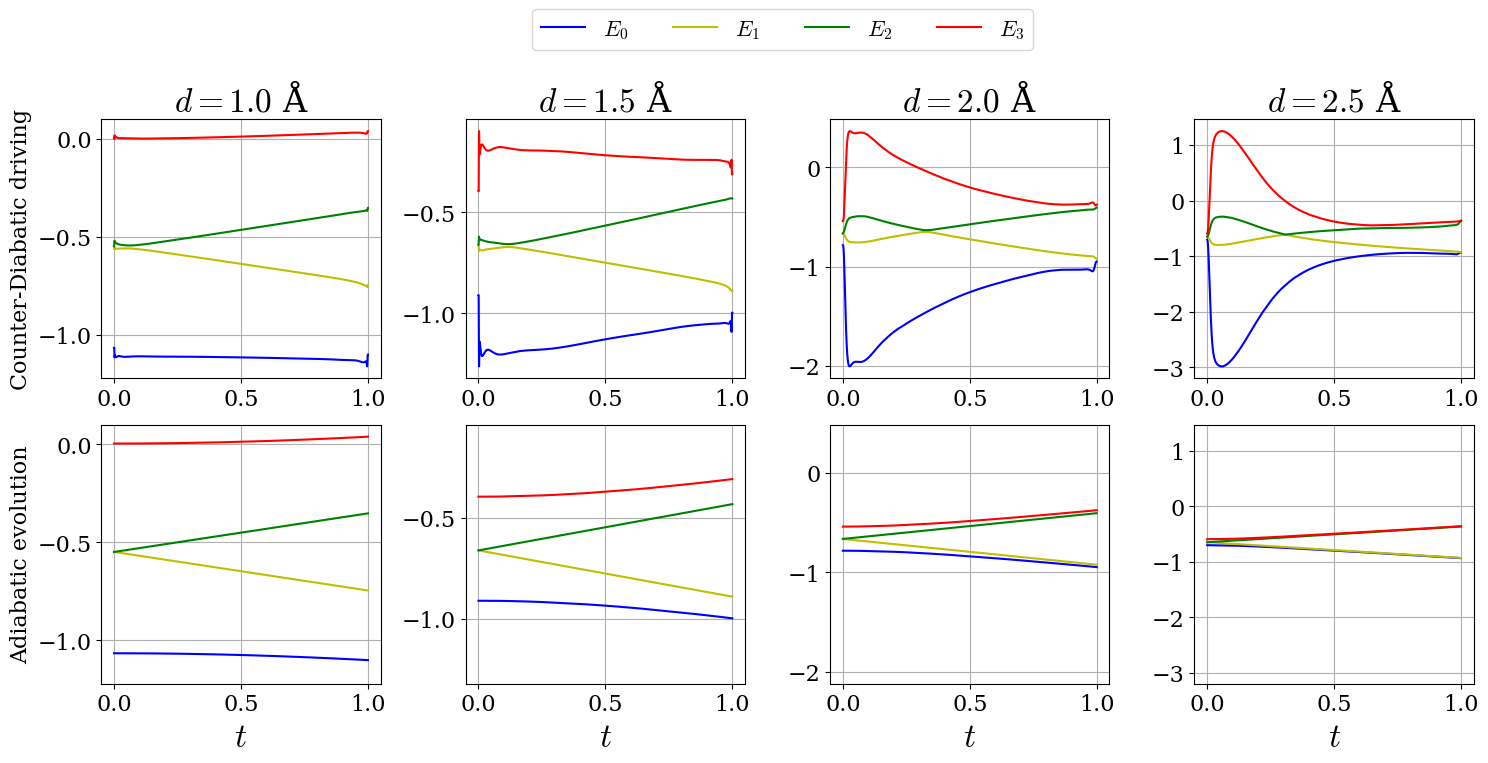} 
\caption{Temporal evolution of the real component of the instantaneous energy levels, namely the eigenvalues $E_{n}(t)$, describing the molecule $\mathrm{H_{2}}$ within a system of 2 qubits utilizing the STO-3G basis. These computations are conducted for diverse values of the interparticle bond distance, $d$. A comparative analysis of the energy levels is presented, showing the results obtained from the CD protocol (\ref{eq:H_PDE}) in the top row, juxtaposed against the levels obtained from the same methodology but with a fully adiabatic transition (\ref{eq:Had}) in the bottom row. It is noteworthy that the energy levels demonstrate a tendency to exhibit greater separation under the CD protocol, a phenomenon that becomes particularly pronounced at $d=2.0\:\text{\AA}$ and $d=2.5\:\text{\AA}$.}
\label{fig:re_eigenvalues}
\end{figure}

\begin{figure}[!htbp]
\centering
\includegraphics[scale=0.42]{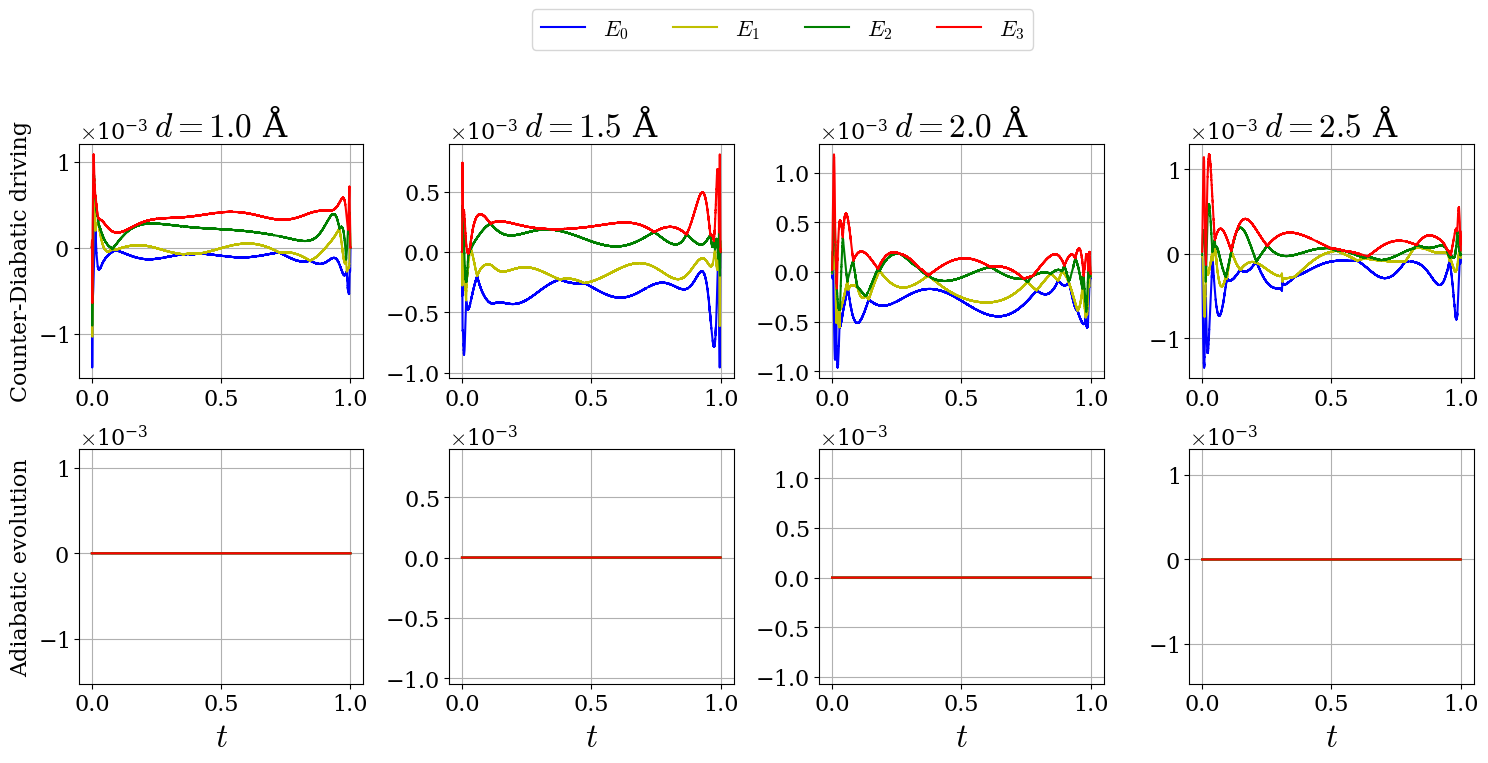} 
\caption{Time-dependent variation of the imaginary component of the eigenvalues $E_{n}(t)$ investigated for the molecular system $\mathrm{H_{2}}$, represented by a 2-qubit configuration in the STO-3G basis. The computational analysis encompasses two scenarios: one employing the CD protocol in the top row, and the other considering a purely adiabatic transition in the bottom row. In the former case, the imaginary components exhibit magnitudes on the order of $10^{-3}$, whereas in the latter case, these are precisely zero, as dictated by the definition of the underlying PDE (\ref{eq:Had}).}
\label{fig:im_eigenvalues}
\end{figure}

The operator $\bm{\mathcal{H}}(t)$, as defined in Equation (\ref{eq:H_PDE}), is specifically designed to yield the eigenstates $\ket{n(t)}$ of $\bm{\mathcal{H}}_{\text{AD}}(t)$ exactly over time. It ensures that no transitions between energy levels, $E_{n}(t)$, are possible \citep{Berry}. This property holds for all eigenstates of $\bm{\mathcal{H}}_{\text{AD}}(t)$, allowing us to interpret the set of states $\ket{n(t)}$ as ``moving eigenstates'' of the total operator $\bm{\mathcal{H}}(t)$. Consequently, we can extract the energy levels corresponding to $\bm{\mathcal{H}}(t)$ throughout the entire time evolution and compare them to the energy levels obtained under the assumption of a completely adiabatic transition, i.e., considering $\frac{d\lambda}{dt}=0$. Figures \ref{fig:re_eigenvalues} and \ref{fig:im_eigenvalues} depict the real and imaginary components, respectively, corresponding to two distinct scenarios considered. The first one employs the CD protocol and is shown in the top row, while the second scenario involves a totally adiabatic transition and is displayed in the bottom row. These investigations were conducted for the $\mathrm{H_{2}}$ molecule. To broaden the scope of our study, we examined different bond distance values between particles, specifically $d\in\{1.0,1.5,2.0,2.5\}\:\text{\AA}$. The numerical values of the corresponding initial ($\bm{\mathcal{H}}_{\text{HF}}$) and final ($\bm{\mathcal{H}}_{\text{problem}}$) Hamiltonian operators are described in Table \ref{Table1}. Notably, Figure \ref{fig:im_eigenvalues} reveals that the imaginary part of the eigenvalues (energies) obtained is on the order of $10^{-3}$, indicating their proximity to zero. As such, these values have negligible influence on observable analyses. However, it is essential to recognize that this outcome directly arises from the hermiticity of the physical observable. Furthermore, by adjusting the weight $\omega_{\text{Coupling}}$, it is possible to further fortify the enforcement of this property. Moreover, in view of the formulation of the Hamiltonian operator under the entirely adiabatic scenario, $\bm{\mathcal{H}}_{\text{AD}}$ (\ref{eq:Had}), and the consideration of the initial and final operators as detailed in Table \ref{Table1}, it is important to note that the scalar nature of the function $\lambda(t)$ ensures the absence of complex numbers. As a result, the imaginary component of the energy levels in the completely adiabatic case is strictly zero, as evidenced in the bottom row of Figure \ref{fig:im_eigenvalues}.

Regarding the real component of the energies, which is of primary interest, it is observed that in the case of a fully adiabatic transition, these energies demonstrate nearly linear temporal evolution, with diminishing separation as particle bonding increases. The close proximity of energy levels leads to an unfavorable outcome wherein transitions between them become more likely for the system. However, such a challenge is addressed in the CD protocol, wherein energy levels tend to be notably more separated throughout the entire evolution domain. This phenomenon becomes more noticeable for the energy ground state, $E_{0}$, and the first excited level, $E_{1}$. Notably, when the bound distance ($d$) is set at $2.0\:\text{\AA}$ and $2.5\:\text{\AA}$, it is evident that these two energy levels remain substantially distant, especially at the initial stages of the evolution of the system. This occurrence is highly desirable from an experimental perspective, as it aims to minimize the probability of transitions between energy levels, especially between the ground and the first excited state, given that the system is initially prepared in the $E_{0}$ level.

\FloatBarrier

\FloatBarrier

\subsection{\texorpdfstring{$\bm{\mathcal{A}}_{\text{CD}}(t)$}{A} operator and its decomposition}

\begin{figure}[!htbp]
\centering
\includegraphics[scale=0.42]{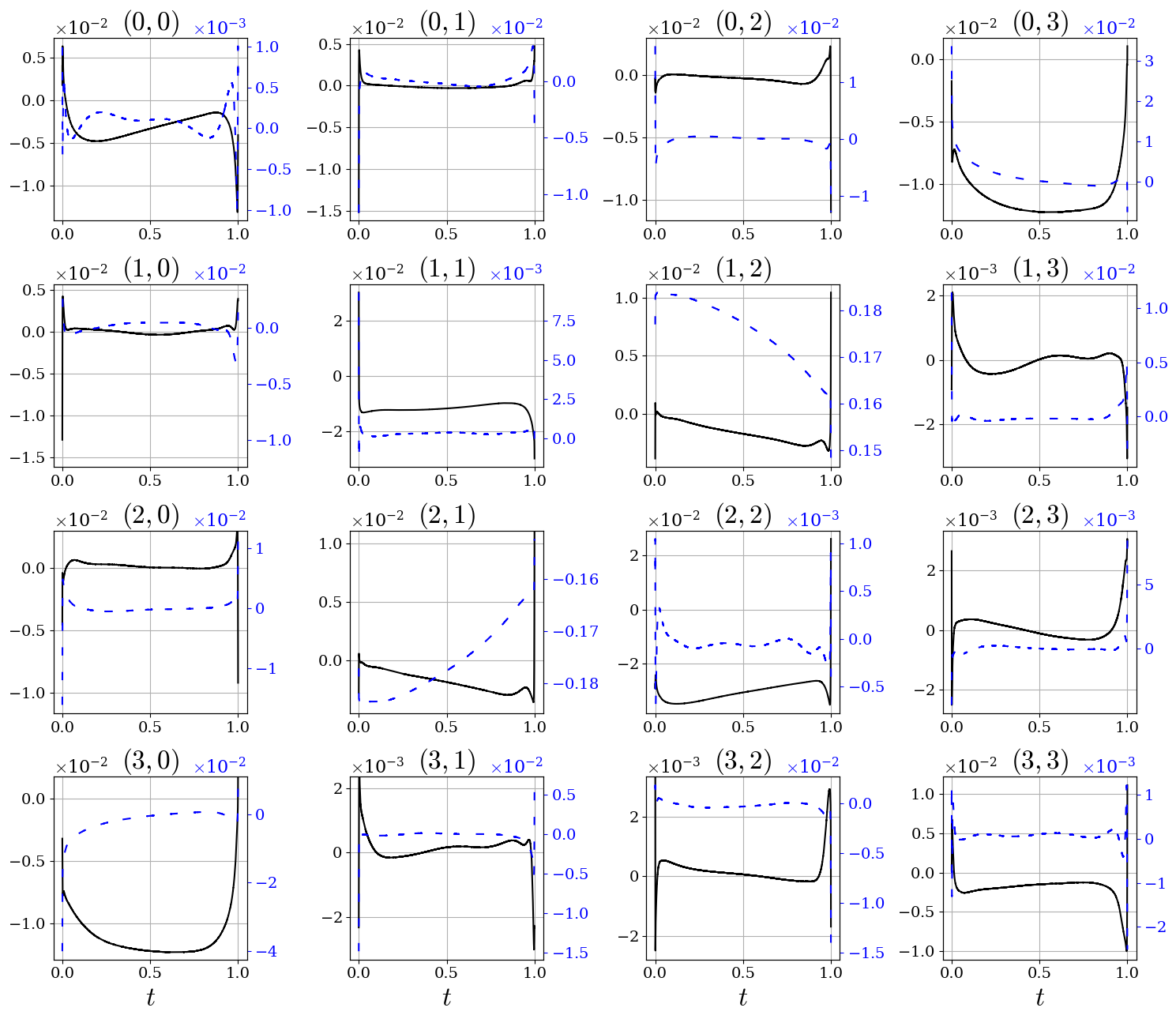} 
\caption{Time evolution of the real and imaginary components of the gauge potential $\bm{\mathcal{A}}_{\text{CD}}(t)$ for the $\mathrm{H_{2}}$ molecule using the STO-3G basis and a bond distance of $d=1.0\:\text{\AA}$, represented respectively in black and blue colors using two different scales. It is observed that the values exhibit variations across different orders of magnitude. Notably, the natural symmetry and antisymmetry arises naturally in these components over time due to the hermeticity of the operator.}
\label{fig:Total_A_CD}
\end{figure}

Our methodology enables us to directly obtain the components of the potential gauge, denoted as $\bm{\mathcal{A}}_{\text{CD}}(t)$. Consequently, we are capable of visualizing the temporal evolution of both its real and imaginary components. Analogous to the presentation of $\bm{\mathcal{H}}(t)$ in Figure \ref{fig:Total_H} above, we display the temporal evolution of the corresponding operator $\bm{\mathcal{A}}_{\text{CD}}$ in Figure \ref{fig:Total_A_CD} of this section. We differentiate their respective real and imaginary components on two distinct scales. Observing the plot, it becomes evident that as a physical observable from an experimental standpoint, the real part of the operator, $\text{Re}\left(\bm{\mathcal{A}}_{\text{CD}}\right)$, exhibits complete symmetry, while its imaginary part, $\text{Im}\left(\bm{\mathcal{A}}_{\text{CD}}\right)$, is fully antisymmetric. This property comes directly from the natural hermiticity of the operator, which, in turn, is imposed indirectly by the condition $\mathcal{L}_{\text{Coupling}}$ (\ref{eq16}), together with $\bm{\mathcal{C}}(t)\in\mathbb{R}^{4^{N{Q}}}$.

While exploring the components of $\bm{\mathcal{A}}_{\text{CD}}(t)$ holds theoretical significance, our primary focus from an experimental point of view lies in obtaining the set of coefficients $\bm{\mathcal{C}}(t)$ over time. These coefficients play a crucial role as they enable the gauge potential to be expressed as a linear combination of all the possible combinations and interactions between the qubits of the system, thereby allowing a better implementation in a real quantum circuit \citep{Digitized-counterdiabatic,Digitized-counterdiabatic_2}. The theoretical formulation of this decomposition is represented by Equation (\ref{eq15}), wherein the potential is expressed using the required Kronecker products. This formulation takes into account all possible combinations for the 2-qubit system under current investigation. Given the relatively small size of our example system, it remains feasible to consider all the possible combinations of the Pauli matrices as the number of these combinations scales with $4^{N_{Q}}$, being $N_{Q}$ the number of qubits. Nevertheless, in certain experimental scenarios, it may be unnecessary to explore all combinations and interactions. In such cases, specific specializations and additional requirements can be applied, guided by the methodology we present.

\begin{figure}[!htbp]
\centering
\includegraphics[scale=0.40]{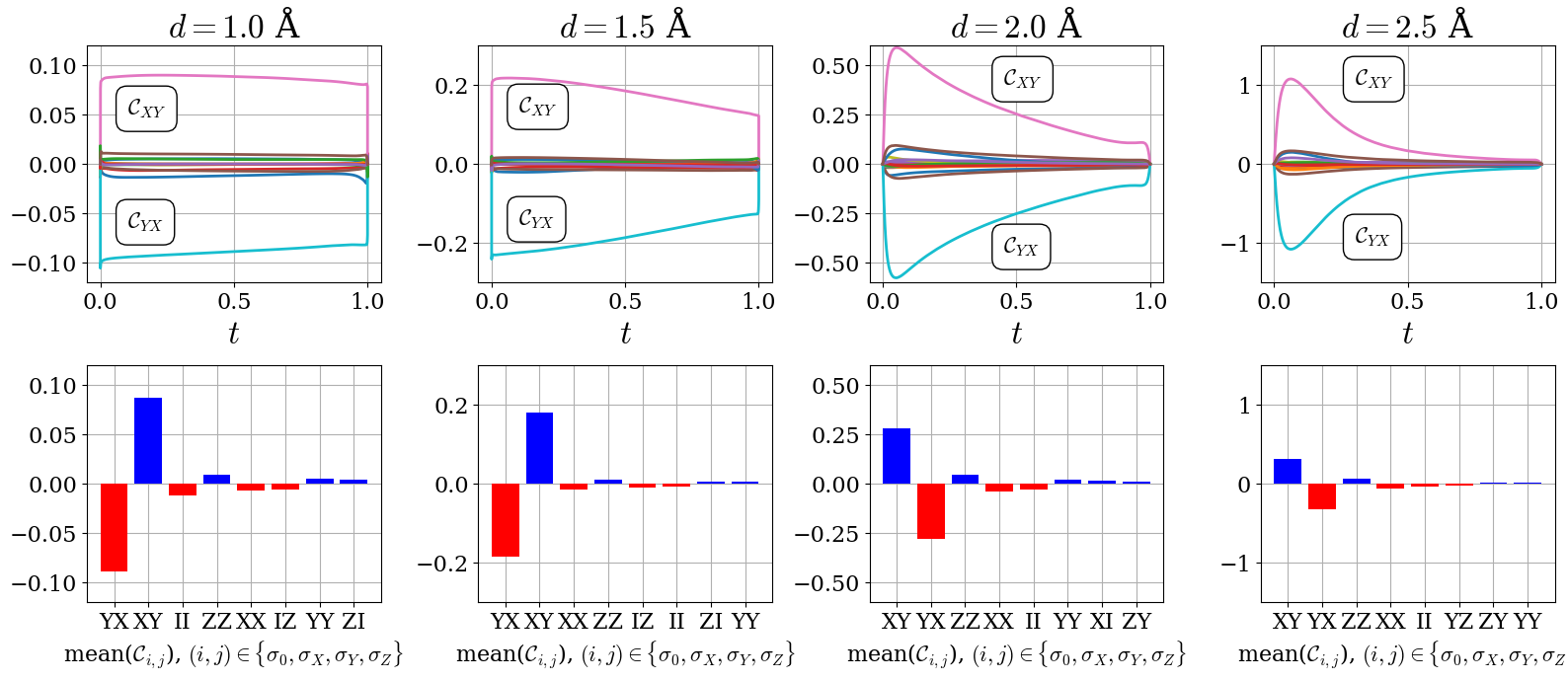} 
\caption{In the upper row, we present the temporal evolutions of the coefficients $\bm{\mathcal{C}}(t)$ resulting from the decomposition of the operator $\bm{\mathcal{A}}_{\text{CD}}(t)$ (\ref{eq15}) applied to the $\mathrm{H_{2}}$ molecule utilizing a 2-qubit configuration in the STO-3G basis. In the lower row, a bar chart illustrates the average values of each coefficient, arranged in descending order of magnitude. It is evident that both the coefficient $\mathcal{C}_{\text{XY}}$ and its symmetric counterpart $\mathcal{C}_{\text{YX}}$ exert the most substantial influence throughout the entire process, followed by $\mathcal{C}_{\text{II}}$, $\bm{\mathcal{C}}_{\text{ZZ}}$ and $\mathcal{C}_{\text{XX}}$.}
\label{fig:C}
\end{figure}

In Figure \ref{fig:C}, we present the temporal evolution of the coefficients derived from the decomposition of the $H2$ system on the STO-3G basis, as a function of the bond distance ($d$) between the particles. The upper row illustrates the evolution itself, while the lower row displays a bar chart presenting the specific coefficients arranged in descending order based on their average values throughout the observed time interval.

This visualization enables us to identify the most significant contributions in terms of the absolute value when implementing this system in a real quantum circuit. Notably, the two coefficients that exhibit the highest values are denoted as $\mathcal{C}_{\text{XY}}$ and its symmetric counterpart $\mathcal{C}_{\text{YX}}$. These findings align with previous literature that employed the NC methodology \citep{Digitized-counterdiabatic, Digitized-counterdiabatic_2}. Our approach naturally reveals these prominent contributions, facilitating an explicit understanding of their respective orders of magnitude. Additionally, there are less prominent contributions, such as $\mathcal{C}_{\text{ZZ}}$, $\mathcal{C}_{\text{XX}}$, and $\mathcal{C}_{\text{II}}$, which warrant consideration as well.

\FloatBarrier

The determination of these coefficients is universally applicable, i.e., it emerges as an outcome of implementing our methodology for any system, irrespective of the number of qubits considered. Nevertheless, as previously commented in the preceding section, certain specific instances may arise in experimental scenarios where it becomes impractical to account for all possible contributions in the decomposition of $\bm{\mathcal{A}}_{\text{CD}} (t)$. In such circumstances, it would be interesting to adapt the method and restrict the output $\bm{\mathcal{C}}(t)$ of the neural network to a subset of the entire set, which becomes particularly relevant when dealing with an increasing number of qubits. Consequently, a trade-off between purely theoretical outcomes, exemplified herein, and those of particular experimental significance may always exist.

\FloatBarrier

\subsection{Scalability}
\label{sec:scalability}

So far, we have conducted a study on a 2-qubit system representing the $\mathrm{H_{2}}$ molecule in the STO-3G basis. However, it is straightforward and easily feasible to modify the base or adjust the number of qubits used in our methodology. This process primarily involves examining the matrix dimensions of the respective network outputs. Specifically, the matrices $\bm{\mathcal{H}},\bm{\mathcal{A}}_{\text{CD}},\bm{\mathcal{H}}_{\text{HF}},\bm{\mathcal{H}}_{\text{problem}}$ are elements of the matrix space $\mathcal{M}_{2^{N_{Q}}\times 2^{N_{Q}}}(\mathbb{C})$, while $\bm{\mathcal{C}}(t)$ belongs to $\mathbb{R}^{4^{N_{Q}}}$. Consequently, the number of components of the variables obtained may increase with the number of qubits, but it is important to note that the neural architecture for all calculations has consistently comprised six internal layers, each containing 30 neurons. This architectural choice is widely documented in the literature and has resulted in state-of-the-art outcomes across various domains of research \citep{Zhiping_2020,GA_PINNs,Jeremy_2022}.

In general, the scale and arrangement of the neural architecture of the PINN will depend on the specific problem being addressed, meaning that it is more or less directly related to the difficulty presented by the underlying PDEs. In our case, we are dealing with a single differential equation, which essentially defines $\bm{\mathcal{H}}(t)$ in Equation (\ref{eq:H_PDE}) whereby involving only one derivative function. Consequently, our chosen neural network architecture efficiently addresses the problem of the CD protocol, as increasing the number of trainable weights usually does not get translated into better results. Moreover, such an increase could even negatively impact computation time and model convergence during the backpropagation process \citep{Panos_2023}. Apart from the $\mathrm{H_{2}}$ molecule, others can also be considered on the STO-3G basis such as lithium hydride, $\mathrm{LiH}$, which can be represented using 4 qubits. The initial and final Hamiltonian operators of the process can be computed again using \citep{Qiskit}.

\begin{figure}[!htbp]
\centering
\includegraphics[scale=0.50]{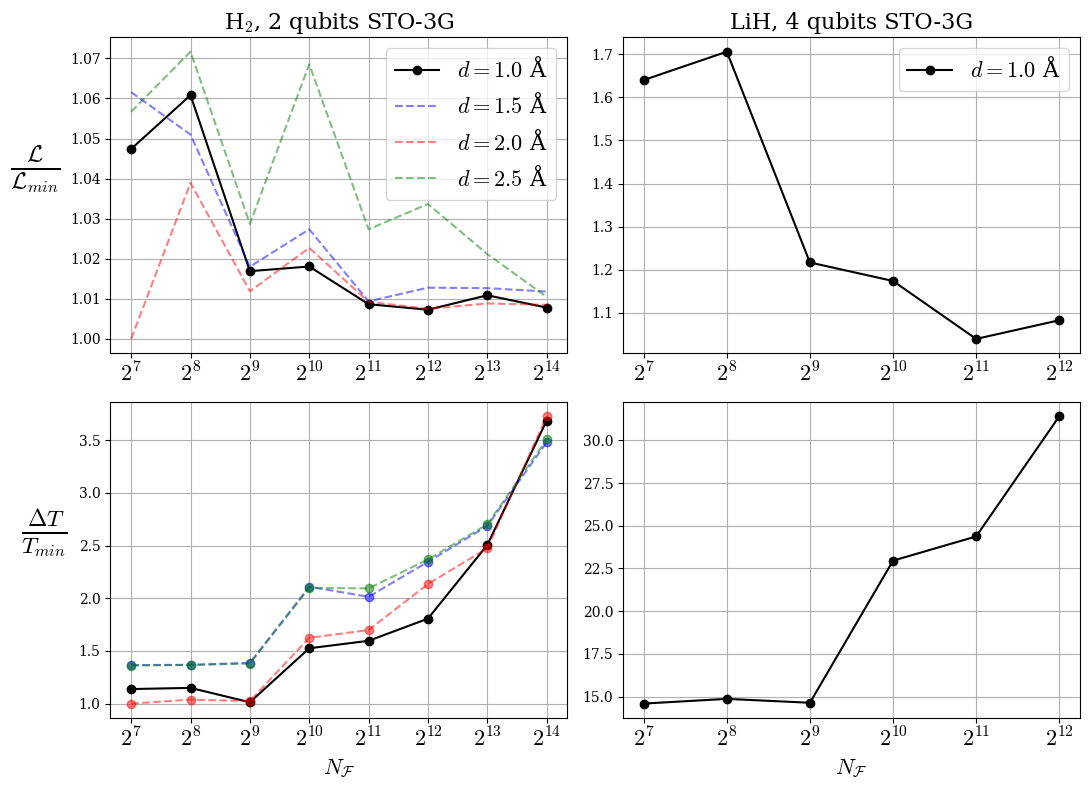} 
\caption{Graphical investigation of the scalability of our methodology for the $\mathrm{H_{2}}$ molecule in the STO-3G basis. Various bond distances of the hydrogen atoms are considered, represented by distinct colors as indicated in the legend to the right. The top side of the graph illustrates the total physical loss $\mathcal{L}$ (\ref{eq18}) after completing the training process, plotted against the number of points $N_{\mathcal{F}}$ considered within the domain $t\in\left(t_{\text{min}},t_{\text{max}}\right)$. On the bottom side of the graph, we present the time required to complete the entire training process, normalized to the minimum time. We conducted these experiments using an NVIDIA A16 GPU.}
\label{fig:Scalability}
\end{figure}

Discussing scalability within these methodologies holds substantial importance as it elucidates the extent to which an approach can be experimentally applicable. Undoubtedly, our DL-based methodology provides a wealth of information, as we have shown throughout the entire paper. However, when addressing the scalability of the method, we need an examination of two key factors: the number of qubits and the quantity of points encompassed within the time interval ($N_{\mathcal{F}}$), in conjunction with the graphics processing unit (GPU) or hardware employed, as a whole. All computations considered employed an NVIDIA A16 card with a memory capacity of 16 GB. Consequently, we present, in Figure \ref{fig:Scalability} (top row), a comprehensive analysis of the final physical loss (or final total residual) which marks the conclusion of network training concerning the number of points within the interval $\left (t_{\text{min}},t_{\text{max}}\right)$ denoted as $N_{\mathcal{F}}$. This analysis encompasses diverse bond distances for the $\mathrm{H_{2}}$ molecule (left) and a singular value of $d$ for the $\mathrm{LiH}$ molecule (right), solely for the purpose of facilitating comparisons.

\FloatBarrier

The physical loss depicted in the top row of the figure has been normalized with respect to the minimum value observed for that particular magnitude across all training sessions considered in this section, encompassing both the $\mathrm{H_{2}}$ and $\mathrm{LiH}$ simulations. This normalization, denoted as $\mathcal{L}_{\text{min}}$, allows us to represent the loss in a manner independent of specific training instances. An analysis of the results reveals that the general loss for the $\mathrm{LiH}$ case is marginally higher, although the values are within the same order of magnitude as the $\mathrm{H_{2}}$ case. It is worth noting that both simulations employed an identical neural architecture comprising 6 internal layers, each containing 30 neurons. Furthermore, a common training regimen of 500,000 iterations (epochs) was applied to all cases. Consequently, a longer training duration would likely result in reduced final physical errors, owing to the common training approach. The consistency in architecture and training duration enables us to draw meaningful comparisons between both simulations and enables us to fairly evaluate their respective final performances.

On the other side, for the purpose of effectively quantifying the computational time involved in various training sessions and facilitating a meaningful comparison, it is imperative to consider external factors that could impact the simulations. These factors include the specific GPU utilized, available free memory in the CPU (as some calculations are delegated to the CPU), among others. In the lower section of the figure, we present a comparative analysis between the two molecules concerning the time consumed ($\Delta T$), which is also normalized with respect to the minimum value obtained for this metric, denoted as $T_{\text{min}}$. The results demonstrate that an increase in the number of data points, $N_{\mathcal{F}}$, generally leads to a multiplication of approximately 3.5 times in the compute time consumed, when transitioning from $2^{7}$ to $2^{14}$ data points, for the $\mathrm{H_{2}}$ molecule. However, it is crucial to highlight that with only $N_{\mathcal{F}}=2^{11}$ the final physical error $\mathcal{L}$ obtained is minimal. This observation indicates that augmenting the sampled time domain does not necessarily enhance the performance of the model. In other words, with this particular number of data points in the training domain, the PINN exhibits significant capabilities in making inferences, extrapolating information to new temporal instances, and interpolating. Consequently, the adoption of $2^{11}$ data points implies a multiplicative factor of slightly more than 1.5 in comparison to the minimum elapsed computation time and it can be regarded as a favorable trade-off between training time and performance.

Concerning the training of the $\mathrm{LiH}$ molecule which involves 4 qubits, a significant increase in the required time is observed compared to the case of $\mathrm{H_{2}}$ with 2 qubits. This discrepancy arises due to the consideration that the latter necessitates a decomposition of the potential gauge $\bm{\mathcal{A}}_{\text{CD}}(t)$ comprising 16 possible combinations, each represented by a $4\times 4$ matrix. However, in the case of the lithium hydride, there are 256 possible combinations in total, each represented by a matrix size of $16\times 16$. Consequently, both hardware memory and training time experience a considerable surge from one case study to another. It is important to note, however, that this increase in resources is essential for extracting all possible information from the problem. This includes the scheduling function, all components of the potential gauge at each time instant, as well as the instantaneous values of each coefficient of the decomposition. In practical situations, theoretical analysis often focuses on a subset of all the possible interactions of the qubits constituting the system. By reducing the number of interactions from 256 to a more manageable figure, the problem becomes more amenable to study. Under these circumstances, the primary contributors to the memory and computing time consumption are both $\mathcal{L}_{\text{Coupling}}$ (\ref{eq16}) and $\mathcal{L}_{\text{Least Action}}$ (\ref{eq13}). The former involves simultaneous manipulation of numerous matrices, while the latter involves performing two commutators. Moreover, both terms span $N_{\mathcal{F}}$ defined points in time, which further adds to the computational complexity.

\section{Discussion and conclusions}
\label{sec:discuss}

In this study we have shown that deep learning methodologies such as Physics-Informed Neural Networks (PINNs) can be used to tackle the problem of counterdiabatic (CD) driving for analog quantum computation, deviating from conventional numerical methodologies \citep{floquet}. Our proposed method is a generalized problem-independent methodology that offers a comprehensive solution, encompassing the determination of counterdiabatic terms and the temporal parametrization which empirically demonstrates that the adiabatic theory holds true. The suggested approach provides an unified and effective way to resolve the underlying physics of the problem while also giving the theoretical Pauli matrix decomposition needed to be implemented experimentally. Furthermore, using the CD approach allows to get a greater separation between the ground state, $E_{0}$, and the first excited level, $E_{1}$, throughout a significant portion of the temporal progression. This is a desirable property from an experimental perspective, as mentioned in Section \ref{subsec:energy_levels}.

Several questions may emerge from these findings. Firstly, an exploration of the computational capacity regarding the maximum qubit count achievable through PINN-based simulation is desirable, i.e., scalability. Secondly, there is an opportunity to enhance our methodology by integrating recent PINN advancements, including the incorporation of causality as an additional constraint and enabling the network to autonomously learn activation functions. Lastly, the restrictions of the theoretical Pauli matrix decomposition to encompass hardware-specific limitations introduce the prospect of improving the operational performance of analog quantum computers within experimental contexts.

Currently, all methodologies formulated within this context have been exclusively applied to systems comprising two and four qubits, as discused in Section \ref{sec:scalability}. Indeed, our study reveals that the training duration of a PINN for a 4-qubit system is approximately 15 times greater than that required for the 2-qubit counterpart. However, it is imperative to acknowledge that, empirically, the possible permutations of gates and qubit interconnections are usually restricted. Hence, despite the potential exponential increase in training time for an N-qubit system, the imposed experimental limitations render it feasible to effectively train the methodology for a substantial quantity of qubits.

Aside from reducing the problem, it is also possible to improve the overall performance of our proposal by forcing it to respect the causality principle \citep{wang2022respecting} further to restrict the time evolution of our Hamiltonian operators. Using this approach is sufficient to achieve substantial improvements in terms of physical accuracy. Therefore, it is feasible to reduce the number of temporal points used during the training process without altering the performance of the network. Moreover, implementing dynamically trainable activation functions may help improve performance and convergence time \citep{adaptive_functions}.

Furthermore, the physical losses presented in Figure \ref{fig:Losses} are around $10^{-4}$, thus underscoring the imperative to comprehend the attainable precision of a base PINN in the context of CD protocols. Enhancing the presented methodology could entail the imposition of temporal causal prerequisites and the optimization of the neural architecture. The incorporation of mixture coefficients (\ref{eq:mixture_weights}) within the loss function is a predetermined selection based on inductive physical biases, thereby specifying greater weights for the initial and final loss components to steer the progression of the system. Other constituents associated with physical conditions encompass hyperparameters that can be selected through iterative experimentation. The said coefficients are amenable to alteration during the training process, potentially employing techniques such as the Augmented Lagrangian approach \citep{lag_mult}, which adapts them according to the deviation from each physical condition. Consequently, the presented approach offers opportunities for enhancing the achievements and mitigating physical losses, improving, if feasible, the robustness of the model from a physical perspective.

In conclusion, our work has shown that PINNs-based approaches are a promising methodology that can be effectively used to optimize CD protocols for adiabatic quantum computing. However, despite the substantial advancements achieved within this context, it is evident that ample opportunities for enhancement exist. In particular, it would be worth studying the impact of these results in digitized-counterdiabation quantum computing (DCQC) algorithms \cite{hegade2021shortcuts, cadavid2023efficient}. The aforementioned questions stand as open paths for our future research, aiming to evolve and elaborate upon them.

\section*{Acknowledgements}

The authors express their sincere appreciation for the thoughtful attention provided by the Kipu Quantum team. AFS and JDMG are partially supported by the agreement funded by the European Union, between the Valencian Ministry of Innovation, Universities, Science and Digital Society, and the network of research centers in Artificial Intelligence (Valencian Foundation valgrAI). It has also been funded by the Valencian Government grant with reference number CIAICO/2021/184; the Spanish Ministry of Economic Affairs and Digital Transformation through the QUANTUM ENIA project call – Quantum Spain project, and the European Union through the Recovery, Transformation and Resilience Plan – NextGenerationEU within the framework of the Digital Spain 2025 Agenda.

\bibliographystyle{unsrt}
\bibliography{references}

\end{document}